%% file: paper.tex
\documentclass[10pt,conference]{IEEEtran}
\IEEEoverridecommandlockouts
\usepackage{cite}
\usepackage{amsmath,amssymb,amsfonts}
\usepackage{algorithmic}
\usepackage{graphicx}
\usepackage{textcomp}
\usepackage{xcolor}
\def\BibTeX{{\rm B\kern-.05em{\sc i\kern-.025em b}\kern-.08em
    T\kern-.1667em\lower.7ex\hbox{E}\kern-.125emX}}

\usepackage{xspace}
\usepackage[detect-all=true,mode=text]{siunitx}
\usepackage{colortbl}
\usepackage[inline]{enumitem}
\usepackage{booktabs}
\usepackage{multirow}
\usepackage{multicol}
\usepackage{subcaption}
\usepackage{arydshln}
\usepackage{rotating}
\usepackage{tabularray}
\usepackage{tikz}
\usepackage[flushmargin]{footmisc}
\usepackage{layouts}
\usepackage{flushend}
\usepackage{hyperref}

\newcommand{\ie}{i.e.,\xspace}
\newcommand{\eg}{e.g.,\xspace}

\newcommand{\afblock}[1]{\textbf{#1:}}
\newcommand{\takeaway}[1]{\textit{\afblock{Takeaway} #1}}

\newif\ifmanualheader

\newif\ifdraft

\drafttrue
\draftfalse

\newif\ifauthor
\authortrue

\ifdraft
\newcommand{\todo}[1]{{\textcolor{red}{\textbf{TODO:} #1}}}

\newcommand{\md}[1]{{\color{blue}{#1}}}
\newcommand{\mh}[1]{{\color{green}{#1}}}
\newcommand{\jp}[1]{{\color{purple}{#1}}}
\newcommand{\jl}[1]{{\color{brown}{#1}}}

\else
\newcommand{\todo}[1]{}

\newcommand{\md}[1]{{}}
\newcommand{\jp}[1]{{}}
\newcommand{\mh}[1]{{}}
\newcommand{\jl}[1]{{}}
\fi

\ifauthor
    \usepackage{tikz}
    \newcommand\copyrighttext{
        \footnotesize
        \textcopyright 2024 IEEE.
        Personal use of this material is permitted.
        Permission from IEEE must be obtained for all other uses, in any current or future media, including reprinting/republishing this material for advertising or promotional purposes, creating new collective works, for resale or redistribution to servers or lists, or reuse of any copyrighted component of this work in other works.
        DOI: \href{https://doi.org/10.1109/NOMS59830.2024.10574963}{10.1109/NOMS59830.2024.10574963}
    }
    \newcommand\copyrightnotice{%
    \begin{tikzpicture}[remember picture,overlay]
    \node[anchor=south,yshift=25pt] at (current page.south) {\parbox{\dimexpr\textwidth-\fboxsep-\fboxrule\relax}{\copyrighttext}};
    \end{tikzpicture}
    }
    \IEEEoverridecommandlockouts
\fi

\DeclareSIUnit\permille{\text{\textperthousand}}
\DeclareSIUnit{\billionlong}{\text{billion}}
\DeclareSIUnit{\billion}{\text{B}}
\DeclareSIUnit{\million}{\text{M}}
\DeclareSIUnit{\kilo}{\text{k}}

\newcommand{\scannedipvsixs}{\SI{14}{\billionlong}}

\definecolor{perfecthost}{HTML}{481568}
\definecolor{sensiblehost}{HTML}{3E8A8D}
\definecolor{insecurehost}{HTML}{DCE319}

\definecolor{limegreen}{HTML}{32CD32}

\newcommand{\DefineRemark}[2]{%
  \expandafter\newcommand\csname rmk-#1\endcsname{#2}%
}
\newcommand{\Remark}[1]{\csname rmk-#1\endcsname}
\input{variables/overview.tex}
\input{variables/textvalues.tex}
\input{variables/addressorigin.tex}
\input{variables/security.tex}

\begin{document}

\title{Unconsidered Installations:\\Discovering IoT Deployments in the IPv6 Internet}

\author{
\IEEEauthorblockN{Markus Dahlmanns\(^*\), Felix Heidenreich\(^*\), Johannes Lohmöller\(^*\),\\ Jan Pennekamp\(^*\), Klaus Wehrle\(^*\), and Martin Henze\(^{\S,\ddagger}\)}
\IEEEauthorblockA{
  \(^*\)\textit{Communication and Distributed Systems}, RWTH Aachen University, Germany\\
  \(^\S\)\textit{Security and Privacy in Industrial Cooperation}, RWTH Aachen University, Germany\\
  \(^\ddagger\)\textit{Cyber Analysis \& Defense}, Fraunhofer FKIE, Germany\\
\{dahlmanns, heidenreich, lohmoeller, pennekamp, wehrle\}@comsys.rwth-aachen.de \(\cdot\)
henze@cs.rwth-aachen.de}
}

\maketitle

\begin{abstract}
Internet-wide studies provide extremely valuable insight into how operators manage their Internet of Things~(IoT) deployments in reality and often reveal grievances, e.g., significant security issues.
However, while IoT devices often use IPv6, past studies resorted to comprehensively scan the IPv4 address space.
To fully understand how the IoT and all its services and devices is operated, including IPv6-reachable deployments is inevitable---although scanning the entire IPv6 address space is infeasible.

In this paper, we close this gap and examine how to discover IPv6-reachable IoT deployments.
Using three sources of active IPv6 addresses and eleven address generators, we discovered \SI{\Remark{ipv6totalsuccess}}{}~IoT deployments.
We derive that the available address sources are a good starting point for finding IoT deployments.
Additionally, we show that using two address generators is sufficient to cover most found deployments. %
Assessing the security of the deployments, we surprisingly find similar issues as in the IPv4 Internet, although IPv6 deployments might be newer and generally more up-to-date:
Only \SI{\Remark{ipv6pctmissaccesscontrol}}{\percent}~of deployments have access control in place and only \SI{\Remark{ipv6pctvalidtls}}{\percent}~make use of TLS inviting attackers, e.g., to eavesdrop sensitive data.
\end{abstract}

\begin{IEEEkeywords}
Internet of Things, Internet measurements, IPv6%
\end{IEEEkeywords}

\section{Introduction}
The Internet of Things~(IoT) with all its derivatives, \eg{} the Industrial IoT~(IIoT) or smart homes, offers various benefits to users and society alike~\cite{kassab-iotbenefits-2020, KHAN2020106522}.
To realize these benefits, corresponding deployments often interact with the physical world and have to process and communicate highly sensitive data~\cite{henze-dcam-2017,maggi2018fragility}.
Hence, operators must run their deployments with care, e.g., regarding security measures like access control.
For communication, the growing number of IoT deployments benefits from the significant size of the IPv6 address space since IPv4 addresses are rare, depleted, and thus numerically insufficient for all Internet-connected devices~\cite{piraux-ipv4depletion-2022,oecd-ipv6coverage-2020}.
\ifauthor\copyrightnotice{}\fi%

Internet-wide studies allow an understanding of how operators manage real and existing Internet-facing deployments at scale, e.g.,~\cite{2020-dahlmanns-imc-opcua, durumeric-heartbleed-2014, nawrocki-tlsquic-2022}.
Thus, their results are indispensable for the standardization of network protocols and the derivation of requirements for intelligent network management tools.
While Internet-wide studies in the complete IPv4 address space~($2^{32}$~addresses) finish in a few minutes~\cite{adrian-zippierzmap-2014}, they are not feasible for all $2^{128}$ IPv6 addresses as measurements would require 600~trillion years at the same speed~(\SI{10}{\giga\bit\per\second}).
Thus, fundamentally new approaches are needed to understand how operators manage their IoT services in the wild.

State-of-the-art Internet measurement methods resort to scanning only parts of the IPv6 address space by relying on
\begin{enumerate*}[label=(\roman*)]
\item \textit{hitlists}~\cite{gasser-hitlist-2018,zirngibl-hitlist-2022} of active IPv6 addresses from different sources, \eg{} DNS or email, and
\item numerous \textit{generators}, e.g.,~\cite{yang-6graph-2022,song-det-2022,hou-6scan-2023}, that take seeds, such as hitlists, to produce further IPv6 addresses which might be in use and thus are valuable to scan.
\end{enumerate*}
While these approaches tend to work well finding Web services~\cite{zirngibl-hitlist-2022,steger-generators-2023}, their practicability for IoT services is not well researched.
Past scans only report a comparably low number of IPv6-reachable IoT(-backend) services~\cite{saidi-iotbackend-2022,jose-ipv6-2023} while relying on a single hitlist.
Thus, it is unknown how well generators combined with which seeds can increase the number of discoverable IoT deployments (as for Web services).
However, answering this question is key to properly understand whether operators misconfigure IPv6-reachable IoT deployments similarly to deployments in the IPv4 address space~\cite{2020-dahlmanns-imc-opcua,dahlmanns-2022,srinivasa-iotmisconfiguration-2021}, or whether these deployments profit from their (potentially) more recent installation.

In this paper, we thus address the research gap of understanding the findability, prevalence, and configuration of IoT deployments in the IPv6 address space.
More specifically, we combine eleven open-sourced generators and three seedlists to discover IoT services in the IPv6 address space that use common IoT protocols, i.e., AMQP, MQTT, OPC~UA, and CoAP, determine which combination of address sources works best to find IoT deployments, and exemplarily assess their security configuration in comparison to the IPv4 address space.

\noindent\afblock{Contributions} Our main contributions are as follows.
\begin{itemize}[noitemsep,topsep=0pt,leftmargin=9pt]
  \item We adapt eleven state-of-the-art address generators for IoT-focused scanning on three seedlists (from related work, DNS, and data from IPv4 scans) and find \SI{\Remark{ipv6totalsuccess}}{}~IPv6-reachable IoT deployments.
  \item Tracking the origin of the IPv6~addresses, we derive that using all seedlists for scanning is beneficial.
  Still, two generators suffice to find \SI{\Remark{pcthostsSeedAnddnsAAAAwww6Scan6Graph}}{\percent} of all identified IoT deployments.
  \item We show that the security configuration IPv6-reachable IoT deployments does not differ from their IPv4 counterparts, i.e., they show the same problems such as missing communication security~(\SI{\Remark{totalpctmissingsecurity}}{\percent}) and disabled access control~(\SI{\Remark{totalpctmissingaccesscontrol}}{\percent}).
  \item We open-source our used tools to scan and track IPv6 addresses to support future research~\cite{toolchain}.
\end{itemize}

\section{Related Work \& Background}
\label{sec:relatedwork}
\label{sec:background}

Our measurements of IPv6-reachable IoT services are motivated by approaches guiding scanners in the IPv6 Internet to addresses of interest and related work analyzing the operation of IoT deployments in the IPv4 address space.

\subsection{IPv6-wide Scanning}

While scans for the entire IPv4 space finish within minutes~\cite{adrian-zippierzmap-2014}, completely scanning the larger IPv6 address space is infeasible~\cite{rfc5157}.
Thus, as listed in Table~\ref{tab:approaches}, various approaches attempt to identify IP addresses worth scanning.

\afblock{Hitlists}
Early on, researchers proposed to base IPv6 scanning on resources that contain addresses known to be in use~\cite{rfc5157,rfc7707,borgolte-dnssec-2018}, \eg{} based on passive Internet traffic, or DNS, and evaluated how many active IPv6 addresses respective sources can derive~\cite{gasser-hitlist-2016,fiebig-hitlist-2017}.
To provide a starting point for IPv6 scanning, researchers from TUM~\cite{gasser-hitlist-2018} make their addresses  available in a hitlist.
The public availability already backed numerous works, \eg{} analyzing the deployment of QUIC~\cite{zirngibl-quic-2021} or peripheral devices~\cite{xiang-periphery-2021}.

\afblock{Scanlist Generation}
By analyzing hitlists, Gasser et al.\ revealed that used IP addresses are clustered~\cite{gasser-hitlist-2018}. %
Consequently, generation approaches for the IPv6 address space emerged to enable more targeted scanning~(cf.\ Table~\ref{tab:approaches}).
While passive approaches take a \textit{seedlist} to generate \textit{scanlists}, active strategies scan selected IP addresses to incorporate live network information.
The approaches differ in the techniques used to identify new targets, leading to varying effectiveness. %
Recently, the TUM extended their hitlist with IPs generated by 6GAN, 6Graph, 6Tree, and 6VecLM, but already noted reduced effectiveness for finding Web services compared to their original publication~\cite{zirngibl-hitlist-2022}.

\begin{table}
\scriptsize
\centering
\input{tables/gen-overview.tex}
\caption{%
Numerous approaches try to ease IPv6 scanning.
Approaches in \colorbox{lightgray}{gray} are (openly) available.
}
\label{tab:approaches}
\vspace{-2.5em}
\end{table}

\afblock{Alias Detection}
Previously, single servers were found to answer requests to complete IPv6 subnets, potentially biasing measurements~\cite{gasser-hitlist-2018}.
Thus, researchers leveraged protocol features of SNMPv3~\cite{albakour-snmp-2021}, IPv6 fragmentation~\cite{luckie2013speedtrap,beverly-fragmentation-2013}, ICMP rate limiting~\cite{vermeulen-icmprate-2020}, the TTL from different vantage points~\cite{marder-apple-2020}, unused addresses~\cite{padmanabhan-uav6-2015}, and multi-protocol behavior~\cite{albakour-pushing-2023} to ``dealias'' IPs used in the Internet core only.
For host addresses, researchers either utilize the caching behavior of Path Maximum Transmission Units~(``Too Big Trick'')~\cite{song-det-2022} or the sparsity of IPv6 addressing schemes by checking for similar responses in the surrounding subnet of a single IP address~\cite{gasser-hitlist-2018,foremski-entropyip-2016}.

\takeaway{%
Numerous works target to increase the coverage of IPv6 scans, allowing researchers to use hitlists and address generators in every possible combination.
However, until now, the evaluation of their value heavily focused on Web services.
}

\subsection{Past Efforts of IoT Scanning}
Different Internet scan services, \eg{} Censys~\cite{durumeric2015search, censys}, perform active measurements to collect and share meta-information on reachable (IoT)~deployments~\cite{leverett-shodanclassification-2011,hansson-analysisshodan-2018}.
Such meta-information allows attackers to find deployments that are insufficiently configured~\cite{alami-shodaniot-jordan-2017,zhao-iot-2022,ceron2020online,kiravuo-vulnerabilitiesfinland-2015,genge-shovat-2016}.
Still, these services do not find all Internet-reachable deployments~\cite{barbieri2021assessing}.

For several years now, researchers have overcome this issue by collecting their data on Internet-reachable IoT deployments~\cite{yang-iot-2019,srinivasa-iotmisconfiguration-2021, dahlmanns-2022, maggi2018fragility} using, \eg{} ZMap~\cite{durumeric-zmap-2013}.
For example, these measurements evinced more than \num{300000}~MQTT brokers~\cite{srinivasa-iotmisconfiguration-2021, dahlmanns-2022, maggi2018fragility} and more than \num{1000}~OPC~UA deployments~\cite{2020-dahlmanns-imc-opcua}. %
Security analyses of these deployments reveal that many fail to enable access control or end-to-end security~\cite{dahlmanns-2022, 2020-dahlmanns-imc-opcua,srinivasa-iotmisconfiguration-2021,maggi2018fragility}, thus opening many doors for attackers. %

Strikingly, all of these works focus on the IPv4 address space, \ie{} their assessment does not cover the larger~(and possibly more modern) IPv6 part of the Internet. %
Notably, large scanning services only list a handful of IoT-related IPv6 deployments~\cite{shodan,censys} which were classified in a previous work~\cite{wu-iscc-2022}.
Even initial works searching IoT(-backend) services using a hitlist (but no generators)~\cite{saidi-iotbackend-2022,jose-ipv6-2023} only find a limited number. %
It is thus open whether and which address generators enable a broader view of IoT~deployments in the IPv6 Internet. %

\takeaway{%
Internet-wide studies revealed various peculiarities for IoT deployments but focused on IPv4 so far. %
Which technique works best to find IPv6-reachable IoT deployments remains unclear but is fundamental to gain a full view.
}

\section{I\lowercase{o}T-Focused IP\lowercase{v}6 Scanning}
\label{sec:methodologyanddataset}

We augment prior efforts on IPv6 scanning by performing active measurements to specifically analyze the spread of IoT and IIoT deployments.
To ascertain how well specific address generation techniques and address sources perform, we simultaneously track the origin(s) of each scanned IP address.

\subsection{Methodology}
\label{sec:methodology}

\begin{table*}
\scriptsize
\centering
\setlength{\tabcolsep}{2.65pt}
\input{tables/dataset.tex}
\vspace{-0.5em}
\caption{%
\textit{Left:}~Protocols and their variants (\(^\dagger\)standard / \(^\ddagger\)secure port), scan dates, and number of scanned IP addresses.
\textit{Center:}~Results of our validation process.
\textit{Right:}~AS, certificate, and aliasing information on valid deployments.}
\label{tab:dataset}
\vspace{-2em}
\end{table*}

To understand which measurement strategy works best to get insight into how operators manage IPv6-reachable IoT deployments, our methodology combines seedlists and generators with IoT scanning and its ethical needs.

\subsubsection{Protocol Focus}
We focus on four IoT-related protocols that implement modern communication paradigms and were subject to recent studies in the IPv4 address space~\cite{srinivasa-iotmisconfiguration-2021, dahlmanns-2022, 2020-dahlmanns-imc-opcua}.
Specifically, due to their modernity, our scans cover the two Publish-Subscribe protocols AMQP and MQTT, usually used for IoT-backend infrastructures~\cite{srinivasa-iotmisconfiguration-2021, dahlmanns-2022}, as well as CoAP and OPC~UA as promising representatives for (I)IoT-related protocols implemented by IoT devices.
Since we are also interested in potential different (security)~operation of deployments, we scan for both the standard and TLS-secured~(DTLS in case of CoAP) variant of these protocols~(cf.~Table~\ref{tab:dataset}~(left)).
Our protocol selection significantly influences active generators as well as our scanning since it requires support for the specific transport and application layer protocol via the respective port.

\subsubsection{Seedlist Sources}
To understand which source of IPv6 addresses performs best in this regard, we rely on three primary sources for IPv6 addresses: %
\begin{enumerate*}[label=(\roman*)]
\item the input IP addresses for \textit{TUM hitlist}~\cite{gasser-hitlist-2018, zirngibl-hitlist-2022}, as it is a widely established source for scanning IPv6 addresses,
\item AAAA records behind domains~(with and without subdomain \texttt{www}) included in \textit{DNS Zone files}~\cite{czds:online}, as operators might rely on easy to remember domain names to connect to IoT-related services, and
\item AAAA records behind domains set as \textit{RDNS entry} of addresses from or included in \textit{certificates} gathered during a \textit{previous IPv4 scan} on the respective port~(cf.\ Table~\ref{tab:dataset}~(left)).
\end{enumerate*}

\subsubsection{Scanlist Generation}
Since IPv6 scanlist generation approaches~(cf.\ Section~\ref{sec:relatedwork}) promise to find more deployments in the IPv6 address space, we use them to generate further addresses for our scans.
However, the generated address sets of different approaches usually only overlap rarely~\cite{zirngibl-hitlist-2022} and, until now, which approach performs best to find IoT-related deployments remains unclear.
Thus, we use all open-sourced generation approaches to answer this question. %

\afblock{Generator Configuration}
\label{sec:approachconfig}
The possible configurations of the generators influencing the output are manifold, e.g., the number of input and output addresses or internally used clustering approaches.
However, the (comparably) extensive runtime of the approaches prevents evaluating all possible configurations of all generators while relying on up-to-date seedlists and running the IPv6 scan close after an IPv4 scan to uphold its comparability.
Instead, to keep the runtime of the address generation feasible, we use the best-performing configuration from the respective publication.
Still, the extensive runtime of 6VecLM forces us to further reduce the number of input addresses to \SI{10000}{}.
To additionally save GPU resources we run 6GAN and 6GCVAE on \textit{TUM} and \textit{v4} sources only.

\afblock{Generator Input}
To feed the generators with the selected amount of seed addresses, we randomly sample seedlists whenever they include more IP addresses as required as input.
We run multiple instances of passive generators in parallel on different input samples when the generator's runtime permits and the input list has more entries than randomly selected. %

\afblock{Generator Results}
We list the number of scanned IP addresses in Table~\ref{tab:dataset}~(left), which vary per scan due to addresses internally scanned by active approaches and approaches not allowing to set the number of generated target addresses.

\subsubsection{Scanning IPv6 \& IPv4}
To evaluate whether IoT services run behind the IPv6 addresses from our seedlists and the subsequent generation results, we use \texttt{zmapv6}~\cite{tumzmapv6:online} on ports of our curated list~(Table~\ref{tab:dataset}~(left)).
For our accompanying IPv4 scans, we scan the entire address space relying on \texttt{zmap}~\cite{durumeric-zmap-2013}.

Whenever we find IP addresses with a specific port open, we subsequently use \texttt{zgrab2}~\cite{COMSYSzg99:online} to perform application layer handshakes and retrieve configuration information as well as payload data.
To also find deployments running the (D)TLS protocol variant on the standard port, we further retry establishing a (D)TLS connection when the standard application handshake was unsuccessful.

\subsubsection{Ethical Considerations}
\label{sec:ethics}
Since our measurements affect and concern real, potentially resource-constrained IoT deployments, we must carefully follow established research guidelines~\cite{dittrich_menlo-report_2012}, best practices for Internet-wide measurements~\cite{durumeric-zmap-2013}, and regulations enacted by our institutions. %

\afblock{Implications for IoT Deployments}
First, we ensure that we do not send requests to single IPs too frequently as this might overload IoT deployments.
Here, we must consider two spots in our methodology:
(i)~IP addresses occurring in the output of more than one generator, and
(ii)~IP addresses \texttt{zmapv6} outputs several times due to potential \texttt{SYN ACK} duplicates.
While deduplicating IPv4 addresses is comparably easy~(\texttt{zmap} does it by default), it is not possible to reasonably store information on all IPv6 addresses in memory.
Instead, we include (inverse) Bloom filters for the deduplication of IP addresses. %
Second, to not overload IoT devices with (D)TLS handshakes, we program our IPv4 and IPv6 scanners to wait \SI{15}{\minute} between subsequent handshakes to a single deployment.
For MQTT, we further limit the connection time~(\SI{30}{\minute}) and outgoing traffic~(\SI{10}{\mega\byte}) per host. %

\afblock{Load on the Internet}
Additionally, to not overload any autonomous system, we limit our scans sending max.\ \SI{100}{\kilo}~packets per second and randomly order addresses to scan.
Additionally, we closely cooperate with our Network Operation Center to handle potential incoming abuse requests.

\afblock{Contact Information}
To give information on the purpose, scope, and (expected) impact of our research, we serve a website that informs about our research and opt-out possibilities on the same IP addresses that we utilize for our Internet scans.
We refrain from scanning any ``blocklisted'' IP addresses again.
To date, these blocklisted addresses accumulate to \SI{5.8}{\million} IPv4 addresses and \SI{3.3e30}{}~IPv6 addresses, primarily due to previous scanning activities of our institution.
Furthermore, we set up rDNS records for our scanning IP addresses, embed contact information in our client certificate, and include our contact details in protocol messages whenever supported.

\subsection{Validating Responses}
\label{sec:validation}
After running our scans according to our methodology, we need to validate the results to extract responses that prove the operation of an IoT deployment behind an IP address~\cite{dahlmanns-2022}.
To better understand our validation results, we also compare findings from our IPv6 scans to IPv4.

Table~\ref{tab:dataset}~(center) leads through our three-step validation process.
For all hosts that respond to our connection attempts~(column \textit{Hosts}), we \textbf{\textit{first}} filter systems that respond but do not establish a valid TCP connection or answer with faulty UDP packets, e.g., an invalid length field~(column \textit{Transport}).
While we see similarities across IPv4 and IPv6 scans, fluctuations in the number of answering and filtered IPv6 addresses can be traced to single runs of address generators resulting in many IPs in specific ASes.
For example, running 6Forest for OPC~UA~(port~4843) on v4 results generated more than \SI{1}{\million} IPs in a single AS that all respond but do not establish a valid connection.
This result underpins the importance of carefully selecting address generators and their inputs.

\textbf{\textit{Second}}, we check for deployments that complete a (D)TLS handshake~(column \textit{(D)TLS Success}).
Notably, on IPv6, similar to IPv4, numerous hosts complete a (D)TLS handshake on the port specified for the non-(D)TLS variant of the respective protocol, already indicating that some IoT deployments run (D)TLS-enabled protocol deployments on the standard port.

\textbf{\textit{Last}}, we report on deployments that correctly respond to protocol conformant requests~(column \textit{Valid}).
The large discrepancy between valid IoT-related protocol deployments and successful TCP and (D)TLS handshakes underlines that operators ``hide'' several non-IoT-related services behind ports intended for IoT protocols. %
Additionally, some deployments offer TLS on both ports or optional TLS support by providing an insecure and secure endpoint on the respective ports.
In the following, we count these deployments only once and as TLS-adopting as they otherwise would distort our analysis.
For the IoT-related services under study, we find in total~\SI{\Remark{ipv6totalsuccess}}{}~IPv6-reachable deployments with a strong bias towards protocols usually used for backend services~(Backend:~\SI{\Remark{backendipv6totalsuccess}}{}, Device:~\SI{\Remark{deviceipv6totalsuccess}}{}).
In comparison to IPv4, we find fewer deployments~(IPv4:~\SI{\Remark{ipv4totalsuccess}}{}), indicating both that address generators today do not generate all relevant IP addresses and that probably fewer deployments are reachable via IPv6.
Still, the fraction of services that use (D)TLS to secure communication is low:
Only \SI{\Remark{ipv6pctvalidtls}}{\percent} of IPv6-reachable services implement (D)TLS~(IPv4:~\SI{\Remark{ipv4pctvalidtls}}{\percent}) showing that, although implementing a more modern Internet protocol version, only a few deployments and their operators consider security.

\subsection{Information on Valid Deployments}
\label{sec:deployments}

So far, it is unclear in which AS the IoT deployments are located, how many operators run them, and whether they are subject to aliasing.
However, this knowledge is required to allow a better understanding of their operation and the influences our measurement methodology might have on the results.
Table~\ref{tab:dataset}~(right) lists information on the ASes deployments reside in, common names from received certificates, and the number of IPs that may be subject to aliasing.

\afblock{Accommodating ASes}
Since we found fewer IPv6 deployments, the number of ASes where we found IoT deployments is smaller for IPv6 than for IPv4~(column \textit{ASes-Distinct}).
However, the number of ASes per valid deployment is consistently higher.
While this view could be distorted by IPv6 measurements not covering the entire address space, it still indicates that found IPv6 deployments are more distributed over the Internet than all IPv4 deployments.
Additionally, we found valid IPv6 deployments in ASes where no IPv4 deployment is located~(column \textit{ASes-+IPv6}), showing that IPv4-wide studies indeed did not consider all IoT deployments.

Looking at the AS type deployments reside in~(column \textit{ASes-Types}\footnote{\label{foot:peeringdb}AS type according to PeeringDB.}), IPv6 deployments, especially MQTT brokers, are significantly more prominent in content-related systems, e.g., cloud networks.
Interestingly, seedlists, e.g., from TUM, usually contain significantly more addresses in ISP networks~\cite{steger-generators-2023}.
Thus, this shift indicates that ISP addresses in hitlists might have too short a lifetime, e.g., due to prefix changes when reconnecting, or that operators more likely deploy IPv6-reachable IoT backend services in the cloud.

\afblock{Operator Information \& Aliasing}
Similar to the number of ASes per valid deployment, the number of different Common Names per deployment increases as well~(column \textit{Cert. CNs-Distinct}), indicating that fewer operators run multiple (D)TLS-enabled deployments.
Already this large share of Common Names per deployment suggests that the number of aliased addresses in our dataset can be neglected. %
The aliasing information offered with the TUM hitlist confirms this presumption~(column \textit{Aliasing}).
Only \SI{\Remark{ipv6pctaliased}}{\percent} of discovered IPv6 deployments are marked as aliased.

\takeaway{%
Relying on numerous address sources and generators, we identified \SI{\Remark{ipv6totalsuccess}}{} IPv6-reachable IoT deployments with minor subject to aliasing.
Notably, and already looking at their security, a similar small fraction as in IPv4 implements TLS to secure their communication~(\SI{\Remark{ipv6pctvalidtls}}{\percent} vs. \SI{\Remark{ipv4pctvalidtls}}{\percent}).
}

\section{Tracing Found Deployments}
\label{sec:tracing}

To understand which address sources helped to find the IPv6-reachable IoT deployments and to guide future studies, we trace found deployments through our generation process.

\begin{figure}[!t]
\centering
\includegraphics[width=\linewidth]{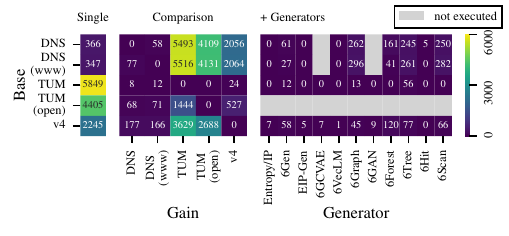}
\vspace{-2.25em}
\caption{Comparison of seedlists and number of generated addresses from generators on specific seedlist leading to previously unknown deployments.}
\label{fig:ip-origin}
\vspace{-1.75em}
\end{figure}

\afblock{Seedlists}
Figure~\ref{fig:ip-origin}~(left) shows how many IPv6 addresses of valid IoT deployments originate from each address source.
Notably, the TUM list leads to the most found deployments and thus constitutes a good starting point for IPv6-wide IoT studies.
However, their openly available hitlist (TUM (open)), where all addresses not answering on Web-specific ports or ICMPv6 are filtered, misses IP addresses running IoT-related services, indicating that checking for an IP address's liveness via these protocols is not beneficial and researchers thus should use the unfiltered list for research offside of the Web.
Interestingly, also IP addresses out of the DNS lead to IoT deployments.
The comparably low gain in finding IoT deployments using DNS in comparison to the TUM list~(Figure~\ref{fig:ip-origin}~(center), \Remark{seedlistgainTUMDNS}~/~\Remark{seedlistgainTUMDNSwww}~(with \texttt{www}) deployments), is due to TUM also including forward DNS data, but of other sources~\cite{gasser-hitlist-2018}.
With \Remark{seedlistgainTUMv4}~additionally found deployments, information from IPv4 scans have still a low, but the comparably highest gain in comparison to the TUM list, showing that IoT-related address sources can increase the value of this list.

\afblock{Generators}
To overcome the narrow view of our selected seedlists, we feed them into address generators.
Figure~\ref{fig:ip-origin}~(right) shows the number of found IoT services that are found due to generator output but are not part of any seedlist.
Most notably, several generators find more active IP addresses using inputs other than the TUM list, indicating that the variety and potentially high age of included addresses do not support current generators.
Instead, using input from DNS leads to the most success in generating IPv6 addresses of IoT deployments as most of these IP addresses are most likely currently in use.

\begin{figure}[!t]
\centering
\includegraphics[width=\linewidth]{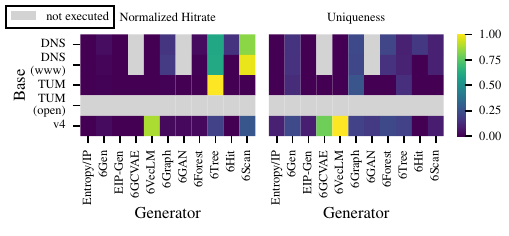}
\vspace{-2.25em}
\caption{Normalized hitrate and uniqueness of generators. Addresses often overlap, but 6Scan has a high hitrate.}
\label{fig:generator-quality}
\vspace{-1.75em}
\end{figure}

To further analyze the effectiveness of the different generators, Figure~\ref{fig:generator-quality}~(left) shows the normalized hitrate, i.e., active IoT-related IP addresses divided by the number of generated IP addresses of each generator, where $1$ corresponds to the generator with the highest hitrate~(unnormalized:~\SI{\Remark{maxhitratepermille}}{\permille}).
While generating the second-highest number of active addresses, 6Scan on the DNS~(www) input has the second-highest hitrate, showing that incorporating information from the Internet during the scan to estimate completely new search directions in addition to the seedlist is beneficial to find IoT deployments.
Thus, 6Scan is a promising address generation candidate for future IPv6-wide IoT studies.

Looking at the generators' result sets, we see intersections.
Figure~\ref{fig:generator-quality}~(right) shows the \textit{uniqueness} of outputs, defined as the number of addresses found by a generator divided by the sum over the number of all generators generating each found address.
Thus, a uniqueness of $1$ means that all addresses are only found by this generator.
6GCVAE and 6VecLM with v4 input have a high uniqueness but only helped to find \Remark{generatorvaliddeploymentsv46GCVAE}~/~\Remark{generatorvaliddeploymentsv46VecLM} IoT deployments, making them unique but not effective.

Our results show that generators help to find \SI{\Remark{totalhostsgeneratorsonly}}{} more IoT deployments than initially included in all seedlists~(\SI{\Remark{totalhostsseedsonly}}{}).
However, running all generators for IoT-related studies is ineffective due to underperformance, overlapping, and long runtimes.
Focussing on a few with comparably high hitrates, i.e., 6Scan and 6Graph on DNS~(www), and all available hitlists allows covering \SI{\Remark{pcthostsSeedAnddnsAAAAwww6Scan6Graph}}{\percent} of all our found deployments only with a fraction of time and computing resources.

\takeaway{%
While the TUM address list is a good starting point for IPv6-wide IoT studies, generators help to find further active addresses.
However, running all generators on all inputs is not required, as their results often overlap. %
}

\section{Security Assessment}
\label{sec:security}

Since a \emph{secure} operation is important for IoT deployments as they regularly get in contact with sensitive (user) data, we exemplarily assess the security of IPv6-reachable IoT deployments in comparison to IPv4.
In the course of our assessment, we adhere to previous assessment approaches~\cite{dahlmanns-2022} to maintain further comparability.
Specifically, we check for the acceptance of (insecure) (D)TLS cipher suites and try to access openly available data.
As we found only a few IPv6-reachable CoAP and OPC~UA deployments, we focus our assessment on AMQP and MQTT deployments.

\begin{figure}[!t]
\centering
\includegraphics[width=\linewidth]{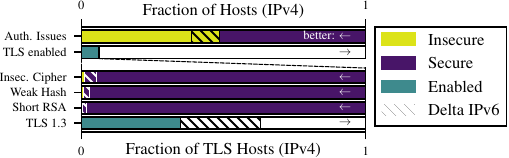}
\vspace{-1.5em}
\caption{Security issues of IPv6 deployments in comparison to IPv4. IPv6 installations have similar issues.}
\label{fig:security}
\vspace{-1.75em}
\end{figure}

\afblock{Access Control}
Figure~\ref{fig:security}~(top) shows that fewer of the found IPv6-reachable deployments miss access control in comparison to IPv4~(\SI{\Remark{ipv6pctmissaccesscontrol}}{\percent} vs.\ \SI{\Remark{ipv4pctmissaccesscontrol}}{\percent}).
However, the focus of found IPv6 deployments running at cloud providers might bias our result, as these potentially have more security expertise than private users.
Still, we found \num{\Remark{mqttcntipv6missaccesscontrol}}~MQTT and \num{\Remark{amqpcntipv6missaccesscontrol}}~AMQP IPv6 deployments that allow anybody to connect.
This negligence might enable attackers to eavesdrop on transmitted IoT data, e.g., from smart homes, or to send malicious commands.

\afblock{Communication Security}
Looking at the \SI{\Remark{ipv6pctvalidtls}}{\percent} of found IPv6-reachable deployments that use TLS to prevent attacks on communications~(\SI{\Remark{ipv4pctvalidtls}}{\percent} for IPv4), a larger share than in the IPv4 address space fail to adhere to TLS configuration guidelines~\cite{nistsp800-tls,rfc7525,tr2102-2}~(\SI{\Remark{ipv6pcttlsmisconfigured}}{\percent} vs.\ \SI{\Remark{ipv4pcttlsmisconfigured}}{\percent}).
Figure~\ref{fig:security}~(bottom) shows that a larger share of IPv6-reachable deployments accepts more insecure ciphers and relies on deprecated primitives, e.g., SHA1, or too short RSA keys in their certificates than IPv4~deployments.
Thus, attackers can more easily impersonate these deployments to access sensitive data or take over IoT systems.

Considering the newest TLS version, i.e., TLS~1.3, we see a larger fraction of deployments signaling support in comparison to IPv4~(\SI{\Remark{ipv6pcttls13}}{\percent} vs.\ \SI{\Remark{ipv4pcttls13}}{\percent}).
This shift suggests that IPv6-reachable deployments are newer and thus can rely on more recent protocol implementations, despite having a penchant for deprecated ciphers and cryptographic primitives.

\takeaway{%
Despite the higher adoption of TLS~1.3 at IPv6 IoT systems, suggesting their more recent deployment~(\SI{\Remark{ipv6pcttls13}}{\percent} vs.\ \SI{\Remark{ipv4pcttls13}}{\percent} in IPv4), they suffer from similar security issues as their IPv4 counterparts, e.g., no access control~(\SI{\Remark{ipv6pctmissaccesscontrol}}{\percent} vs.\ \SI{\Remark{ipv4pctmissaccesscontrol}}{\percent}). %
}

\section{Discussion \& Future Work}
\label{sec:discussion}

The outcome of our work is manifold.
While it supports future IoT and other studies with insight into how to scan IPv6, future work could also address its limitations.

\afblock{Not Covering the Complete Address Space}
All IPv6-wide studies, including ours, cannot reliably estimate which portion of installations they covered.
Still, by scanning \scannedipvsixs{}~addresses, we found \SI{\Remark{ipv6totalsuccess}}{}~IoT deployments helping to understand an even larger part of the IoT.

For IPv4 scans, depending on the research question, it might be sufficient to randomly scan only \SI{1}{\percent} of the address space~\cite{rueth-tcpiw1pct-2019}.
However, in the IPv6 Internet, the sparse usage of addresses would lead to skewed results.
Instead, it might be beneficial to rely on information gathered from NTP pool servers~\cite{rye-ipv6ntp-2023} to increase the low scan-success rate of IPv6 scans~(\SI{\Remark{ipv6permillescansuccess}}{\permille} vs. \SI{\Remark{ipv4permillescansuccess}}{\permille} in IPv4).
Additionally, related work can look into the timeliness of IP addresses in seedlists as other works indicate to find more deployments~\cite{jose-ipv6-2023}. %

\afblock{Address Generator Inputs}
Address generators influence the number of found deployments by nudging the scan to specific addresses.
Thus, their seeding and configuration might influence the study.
While we chose to configure the generators according to the corresponding publications~(cf.\ Section~\ref{sec:approachconfig}) and seed them randomly with IPs separated by source, other approaches could produce more active IPv6 addresses.
Thus, future work still could look into other seedings, e.g., separated by AS type as done for the Web~\cite{steger-generators-2023}.

\afblock{Intersecting Installations}
While we argue that future IoT studies should include IPv6-reachable deployments to gain a full view, our research leaves open how many IoT deployments are reachable via both IPv6 and IPv4. %
Especially given that we found \SI{\Remark{seedlistgainv4v4}}{}~deployments using data from our IPv4 scans as source, the intersection could be high.
Still, we found \SI{\Remark{ipv6deploymentsinipv6onlyases}}{}~IoT deployments located in ASes that do not accommodate any deployment during our IPv4 scan and thus are, with a high probability, not reachable via IPv4.
Future studies could improve the detection of multi-homed deployments by generating fingerprints without considering the IP address.

\section{Conclusion}
\label{sec:conclusion}

Internet-wide studies are an indispensable tool to understand how operators manage IoT deployments in the wild, to uncover security flaws, and to derive requirements for mechanisms preventing misconfigurations in the future~\cite{dahlmanns-2022,srinivasa-iotmisconfiguration-2021}.
However, so far, these studies focused on IPv4-reachable deployments and left out the huge IPv6 address space.

Our results show that not all address generators are beneficial and the seed selection notably influences their result.
Two generators and three address lists suffice to detect \SI{\Remark{pcthostsSeedAnddnsAAAAwww6Scan6Graph}}{\percent} of found deployments.
Security-wise, we find similar issues in the IPv4 and IPv6 address space:
Only \SI{\Remark{ipv6pctvalidtls}}{\percent} of IPv6-reachable deployments implement TLS for communication security~(IPv4:~\SI{\Remark{ipv4pctvalidtls}}{\percent}) and \SI{\Remark{ipv6pctmissaccesscontrol}}{\percent} fail to implement access control~(IPv4:~\SI{\Remark{ipv4pctmissaccesscontrol}}{\percent}), enabling attackers to easily access potentially sensitive information.

To conclude, our work shows that the selection of address generators and their seeding is key to extend the findability of IoT devices in the IPv6 address space beyond hitlists.
Furthermore, mechanisms targeting to prevent insecure misconfigurations and security assessments must also consider IPv6-reachable deployments.

\section*{Acknowledgement}
Funded by the Deutsche Forschungsgemeinschaft (DFG, German Research Foundation) under Germany's Excellence Strategy --- EXC-2023 Internet of Production --- 390621612.

\bibliographystyle{IEEEtran}
\bibliography{paper}

\end{document}

%% file: variables/overview.tex
\DefineRemark{18831883ipv4mqttdate}{05-28}
\DefineRemark{18831883ipv4mqttpctconnsuccess}{19}
\DefineRemark{18831883ipv4mqttpcthostvalid}{0.1}
\DefineRemark{18831883ipv4mqttpcttlsvalidoftransport}{0}
\DefineRemark{18831883ipv4mqttpctvalidtls}{3.6}
\DefineRemark{18831883ipv4mqtttotalas}{3967}
\DefineRemark{18831883ipv4mqtttotalasv6only}{0}
\DefineRemark{18831883ipv4mqtttotalcertcns}{0}
\DefineRemark{18831883ipv4mqtttotalcerts}{0}
\DefineRemark{18831883ipv4mqtttotalconnerr}{3101464}
\DefineRemark{18831883ipv4mqtttotalconnsuccess}{720817}
\DefineRemark{18831883ipv4mqtttotalhosts}{3822281}
\DefineRemark{18831883ipv4mqtttotalips}{3696506441}
\DefineRemark{18831883ipv4mqtttotalsuccess}{359970}
\DefineRemark{18831883ipv4mqtttotalsystemid}{0}
\DefineRemark{18831883ipv4mqtttotaltls}{0}
\DefineRemark{18831883ipv4mqtttotaltlsauth}{0}
\DefineRemark{18831883ipv4mqtttotaltlssuccess}{0}
\DefineRemark{18831883ipv4tlsmqttdate}{05-28}
\DefineRemark{18831883ipv4tlsmqttpctconnsuccess}{19}
\DefineRemark{18831883ipv4tlsmqttpcthostvalid}{0.1}
\DefineRemark{18831883ipv4tlsmqttpcttlsvalidoftransport}{19}
\DefineRemark{18831883ipv4tlsmqttpctvalidtls}{3.6}
\DefineRemark{18831883ipv4tlsmqtttotalas}{80}
\DefineRemark{18831883ipv4tlsmqtttotalasv6only}{0}
\DefineRemark{18831883ipv4tlsmqtttotalcertcns}{299}
\DefineRemark{18831883ipv4tlsmqtttotalcerts}{317}
\DefineRemark{18831883ipv4tlsmqtttotalconnerr}{3101464}
\DefineRemark{18831883ipv4tlsmqtttotalconnsuccess}{720817}
\DefineRemark{18831883ipv4tlsmqtttotalhosts}{3822281}
\DefineRemark{18831883ipv4tlsmqtttotalips}{3696506441}
\DefineRemark{18831883ipv4tlsmqtttotalsuccess}{390}
\DefineRemark{18831883ipv4tlsmqtttotalsystemid}{0}
\DefineRemark{18831883ipv4tlsmqtttotaltls}{135090}
\DefineRemark{18831883ipv4tlsmqtttotaltlsauth}{135090}
\DefineRemark{18831883ipv4tlsmqtttotaltlssuccess}{135081}
\DefineRemark{18831883ipv6mqttdate}{05-31}
\DefineRemark{18831883ipv6mqttpctconnsuccess}{11}
\DefineRemark{18831883ipv6mqttpcthostvalid}{0}
\DefineRemark{18831883ipv6mqttpcttlsvalidoftransport}{0}
\DefineRemark{18831883ipv6mqttpctvalidtls}{3.6}
\DefineRemark{18831883ipv6mqtttotalas}{208}
\DefineRemark{18831883ipv6mqtttotalasv6only}{28}
\DefineRemark{18831883ipv6mqtttotalcertcns}{0}
\DefineRemark{18831883ipv6mqtttotalcerts}{0}
\DefineRemark{18831883ipv6mqtttotalconnerr}{709511}
\DefineRemark{18831883ipv6mqtttotalconnsuccess}{86772}
\DefineRemark{18831883ipv6mqtttotalhosts}{796283}
\DefineRemark{18831883ipv6mqtttotalips}{1828965301}
\DefineRemark{18831883ipv6mqtttotalsuccess}{1675}
\DefineRemark{18831883ipv6mqtttotalsystemid}{0}
\DefineRemark{18831883ipv6mqtttotaltls}{0}
\DefineRemark{18831883ipv6mqtttotaltlsauth}{0}
\DefineRemark{18831883ipv6mqtttotaltlssuccess}{0}
\DefineRemark{18831883ipv6tlsmqttdate}{05-31}
\DefineRemark{18831883ipv6tlsmqttpctconnsuccess}{11}
\DefineRemark{18831883ipv6tlsmqttpcthostvalid}{0}
\DefineRemark{18831883ipv6tlsmqttpcttlsvalidoftransport}{65}
\DefineRemark{18831883ipv6tlsmqttpctvalidtls}{3.6}
\DefineRemark{18831883ipv6tlsmqtttotalas}{8}
\DefineRemark{18831883ipv6tlsmqtttotalasv6only}{3}
\DefineRemark{18831883ipv6tlsmqtttotalcertcns}{7}
\DefineRemark{18831883ipv6tlsmqtttotalcerts}{7}
\DefineRemark{18831883ipv6tlsmqtttotalconnerr}{709511}
\DefineRemark{18831883ipv6tlsmqtttotalconnsuccess}{86772}
\DefineRemark{18831883ipv6tlsmqtttotalhosts}{796283}
\DefineRemark{18831883ipv6tlsmqtttotalips}{1828965301}
\DefineRemark{18831883ipv6tlsmqtttotalsuccess}{11}
\DefineRemark{18831883ipv6tlsmqtttotalsystemid}{0}
\DefineRemark{18831883ipv6tlsmqtttotaltls}{56188}
\DefineRemark{18831883ipv6tlsmqtttotaltlsauth}{56188}
\DefineRemark{18831883ipv6tlsmqtttotaltlssuccess}{56188}
\DefineRemark{48404840ipv4opcuadate}{06-11}
\DefineRemark{48404840ipv4opcuapctconnsuccess}{0.2}
\DefineRemark{48404840ipv4opcuapcthostvalid}{0.1}
\DefineRemark{48404840ipv4opcuapcttlsvalidoftransport}{0}
\DefineRemark{48404840ipv4opcuapctvalidtls}{0}
\DefineRemark{48404840ipv4opcuatotalas}{491}
\DefineRemark{48404840ipv4opcuatotalasv6only}{0}
\DefineRemark{48404840ipv4opcuatotalcertcns}{0}
\DefineRemark{48404840ipv4opcuatotalcerts}{0}
\DefineRemark{48404840ipv4opcuatotalconnerr}{4075623}
\DefineRemark{48404840ipv4opcuatotalconnsuccess}{10062}
\DefineRemark{48404840ipv4opcuatotalhosts}{4085685}
\DefineRemark{48404840ipv4opcuatotalips}{3696506441}
\DefineRemark{48404840ipv4opcuatotalsuccess}{1740}
\DefineRemark{48404840ipv4opcuatotalsystemid}{0}
\DefineRemark{48404840ipv4opcuatotaltls}{0}
\DefineRemark{48404840ipv4opcuatotaltlsauth}{0}
\DefineRemark{48404840ipv4opcuatotaltlssuccess}{0}
\DefineRemark{48404840ipv4tlsopcuadate}{06-11}
\DefineRemark{48404840ipv4tlsopcuapctconnsuccess}{0.4}
\DefineRemark{48404840ipv4tlsopcuapcthostvalid}{0.1}
\DefineRemark{48404840ipv4tlsopcuapcttlsvalidoftransport}{53}
\DefineRemark{48404840ipv4tlsopcuapctvalidtls}{100}
\DefineRemark{48404840ipv4tlsopcuatotalas}{1}
\DefineRemark{48404840ipv4tlsopcuatotalasv6only}{0}
\DefineRemark{48404840ipv4tlsopcuatotalcertcns}{2}
\DefineRemark{48404840ipv4tlsopcuatotalcerts}{2}
\DefineRemark{48404840ipv4tlsopcuatotalconnerr}{4066329}
\DefineRemark{48404840ipv4tlsopcuatotalconnsuccess}{17616}
\DefineRemark{48404840ipv4tlsopcuatotalhosts}{4083945}
\DefineRemark{48404840ipv4tlsopcuatotalips}{3696506441}
\DefineRemark{48404840ipv4tlsopcuatotalsuccess}{2}
\DefineRemark{48404840ipv4tlsopcuatotalsystemid}{1}
\DefineRemark{48404840ipv4tlsopcuatotaltls}{9311}
\DefineRemark{48404840ipv4tlsopcuatotaltlsauth}{9309}
\DefineRemark{48404840ipv4tlsopcuatotaltlssuccess}{6098}
\DefineRemark{48404840ipv6opcuadate}{06-13}
\DefineRemark{48404840ipv6opcuapctconnsuccess}{0}
\DefineRemark{48404840ipv6opcuapcthostvalid}{0.1}
\DefineRemark{48404840ipv6opcuapcttlsvalidoftransport}{0}
\DefineRemark{48404840ipv6opcuapctvalidtls}{0}
\DefineRemark{48404840ipv6opcuatotalas}{3}
\DefineRemark{48404840ipv6opcuatotalasv6only}{2}
\DefineRemark{48404840ipv6opcuatotalcertcns}{0}
\DefineRemark{48404840ipv6opcuatotalcerts}{0}
\DefineRemark{48404840ipv6opcuatotalconnerr}{2680053}
\DefineRemark{48404840ipv6opcuatotalconnsuccess}{261}
\DefineRemark{48404840ipv6opcuatotalhosts}{2680314}
\DefineRemark{48404840ipv6opcuatotalips}{1849275639}
\DefineRemark{48404840ipv6opcuatotalsuccess}{6}
\DefineRemark{48404840ipv6opcuatotalsystemid}{0}
\DefineRemark{48404840ipv6opcuatotaltls}{0}
\DefineRemark{48404840ipv6opcuatotaltlsauth}{0}
\DefineRemark{48404840ipv6opcuatotaltlssuccess}{0}
\DefineRemark{48404840ipv6tlsopcuadate}{06-13}
\DefineRemark{48404840ipv6tlsopcuapctconnsuccess}{0}
\DefineRemark{48404840ipv6tlsopcuapcthostvalid}{0.1}
\DefineRemark{48404840ipv6tlsopcuapcttlsvalidoftransport}{67}
\DefineRemark{48404840ipv6tlsopcuapctvalidtls}{100}
\DefineRemark{48404840ipv6tlsopcuatotalas}{0}
\DefineRemark{48404840ipv6tlsopcuatotalasv6only}{1}
\DefineRemark{48404840ipv6tlsopcuatotalcertcns}{0}
\DefineRemark{48404840ipv6tlsopcuatotalcerts}{0}
\DefineRemark{48404840ipv6tlsopcuatotalconnerr}{2679577}
\DefineRemark{48404840ipv6tlsopcuatotalconnsuccess}{731}
\DefineRemark{48404840ipv6tlsopcuatotalhosts}{2680308}
\DefineRemark{48404840ipv6tlsopcuatotalips}{1849275639}
\DefineRemark{48404840ipv6tlsopcuatotalsuccess}{0}
\DefineRemark{48404840ipv6tlsopcuatotalsystemid}{0}
\DefineRemark{48404840ipv6tlsopcuatotaltls}{493}
\DefineRemark{48404840ipv6tlsopcuatotaltlsauth}{493}
\DefineRemark{48404840ipv6tlsopcuatotaltlssuccess}{49}
\DefineRemark{48434843ipv4httpdate}{06-10}
\DefineRemark{48434843ipv4httppctconnsuccess}{0.4}
\DefineRemark{48434843ipv4httppcthostvalid}{0.1}
\DefineRemark{48434843ipv4httppcttlsvalidoftransport}{0}
\DefineRemark{48434843ipv4httppctvalidtls}{100}
\DefineRemark{48434843ipv4httptotalas}{0}
\DefineRemark{48434843ipv4httptotalasv6only}{0}
\DefineRemark{48434843ipv4httptotalcertcns}{0}
\DefineRemark{48434843ipv4httptotalcerts}{0}
\DefineRemark{48434843ipv4httptotalconnerr}{5200365}
\DefineRemark{48434843ipv4httptotalconnsuccess}{18604}
\DefineRemark{48434843ipv4httptotalhosts}{5218969}
\DefineRemark{48434843ipv4httptotalips}{3696506441}
\DefineRemark{48434843ipv4httptotalsuccess}{0}
\DefineRemark{48434843ipv4httptotalsystemid}{0}
\DefineRemark{48434843ipv4httptotaltls}{0}
\DefineRemark{48434843ipv4httptotaltlsauth}{0}
\DefineRemark{48434843ipv4httptotaltlssuccess}{0}
\DefineRemark{48434843ipv4tlshttpdate}{06-10}
\DefineRemark{48434843ipv4tlshttppctconnsuccess}{0.4}
\DefineRemark{48434843ipv4tlshttppcthostvalid}{0.1}
\DefineRemark{48434843ipv4tlshttppcttlsvalidoftransport}{52}
\DefineRemark{48434843ipv4tlshttppctvalidtls}{100}
\DefineRemark{48434843ipv4tlshttptotalas}{1}
\DefineRemark{48434843ipv4tlshttptotalasv6only}{0}
\DefineRemark{48434843ipv4tlshttptotalcertcns}{3}
\DefineRemark{48434843ipv4tlshttptotalcerts}{3}
\DefineRemark{48434843ipv4tlshttptotalconnerr}{5200365}
\DefineRemark{48434843ipv4tlshttptotalconnsuccess}{18604}
\DefineRemark{48434843ipv4tlshttptotalhosts}{5218969}
\DefineRemark{48434843ipv4tlshttptotalips}{3696506441}
\DefineRemark{48434843ipv4tlshttptotalsuccess}{3}
\DefineRemark{48434843ipv4tlshttptotalsystemid}{1}
\DefineRemark{48434843ipv4tlshttptotaltls}{9702}
\DefineRemark{48434843ipv4tlshttptotaltlsauth}{9700}
\DefineRemark{48434843ipv4tlshttptotaltlssuccess}{6370}
\DefineRemark{48434843ipv6httpdate}{06-11}
\DefineRemark{48434843ipv6httppctconnsuccess}{0}
\DefineRemark{48434843ipv6httppcthostvalid}{0.2}
\DefineRemark{48434843ipv6httppcttlsvalidoftransport}{0}
\DefineRemark{48434843ipv6httppctvalidtls}{100}
\DefineRemark{48434843ipv6httptotalas}{0}
\DefineRemark{48434843ipv6httptotalasv6only}{20}
\DefineRemark{48434843ipv6httptotalcertcns}{0}
\DefineRemark{48434843ipv6httptotalcerts}{0}
\DefineRemark{48434843ipv6httptotalconnerr}{3561454}
\DefineRemark{48434843ipv6httptotalconnsuccess}{524}
\DefineRemark{48434843ipv6httptotalhosts}{3561978}
\DefineRemark{48434843ipv6httptotalips}{1785421878}
\DefineRemark{48434843ipv6httptotalsuccess}{0}
\DefineRemark{48434843ipv6httptotalsystemid}{0}
\DefineRemark{48434843ipv6httptotaltls}{0}
\DefineRemark{48434843ipv6httptotaltlsauth}{0}
\DefineRemark{48434843ipv6httptotaltlssuccess}{0}
\DefineRemark{48434843ipv6tlshttpdate}{06-11}
\DefineRemark{48434843ipv6tlshttppctconnsuccess}{0}
\DefineRemark{48434843ipv6tlshttppcthostvalid}{0.2}
\DefineRemark{48434843ipv6tlshttppcttlsvalidoftransport}{56}
\DefineRemark{48434843ipv6tlshttppctvalidtls}{100}
\DefineRemark{48434843ipv6tlshttptotalas}{0}
\DefineRemark{48434843ipv6tlshttptotalasv6only}{1}
\DefineRemark{48434843ipv6tlshttptotalcertcns}{0}
\DefineRemark{48434843ipv6tlshttptotalcerts}{0}
\DefineRemark{48434843ipv6tlshttptotalconnerr}{3561454}
\DefineRemark{48434843ipv6tlshttptotalconnsuccess}{524}
\DefineRemark{48434843ipv6tlshttptotalhosts}{3561978}
\DefineRemark{48434843ipv6tlshttptotalips}{1785421878}
\DefineRemark{48434843ipv6tlshttptotalsuccess}{0}
\DefineRemark{48434843ipv6tlshttptotalsystemid}{0}
\DefineRemark{48434843ipv6tlshttptotaltls}{296}
\DefineRemark{48434843ipv6tlshttptotaltlsauth}{296}
\DefineRemark{48434843ipv6tlshttptotaltlssuccess}{69}
\DefineRemark{56715671ipv4amqpdate}{06-03}
\DefineRemark{56715671ipv4amqppctconnsuccess}{4.1}
\DefineRemark{56715671ipv4amqppcthostvalid}{0.1}
\DefineRemark{56715671ipv4amqppcttlsvalidoftransport}{0}
\DefineRemark{56715671ipv4amqppctvalidtls}{21}
\DefineRemark{56715671ipv4amqptotalas}{0}
\DefineRemark{56715671ipv4amqptotalasv6only}{0}
\DefineRemark{56715671ipv4amqptotalcertcns}{0}
\DefineRemark{56715671ipv4amqptotalcerts}{0}
\DefineRemark{56715671ipv4amqptotalconnerr}{5129014}
\DefineRemark{56715671ipv4amqptotalconnsuccess}{216882}
\DefineRemark{56715671ipv4amqptotalhosts}{5345896}
\DefineRemark{56715671ipv4amqptotalips}{3696506441}
\DefineRemark{56715671ipv4amqptotalsuccess}{0}
\DefineRemark{56715671ipv4amqptotalsystemid}{0}
\DefineRemark{56715671ipv4amqptotaltls}{0}
\DefineRemark{56715671ipv4amqptotaltlsauth}{0}
\DefineRemark{56715671ipv4amqptotaltlssuccess}{0}
\DefineRemark{56715671ipv4tlsamqpdate}{06-03}
\DefineRemark{56715671ipv4tlsamqppctconnsuccess}{4.1}
\DefineRemark{56715671ipv4tlsamqppcthostvalid}{0.1}
\DefineRemark{56715671ipv4tlsamqppcttlsvalidoftransport}{95}
\DefineRemark{56715671ipv4tlsamqppctvalidtls}{21}
\DefineRemark{56715671ipv4tlsamqptotalas}{790}
\DefineRemark{56715671ipv4tlsamqptotalasv6only}{0}
\DefineRemark{56715671ipv4tlsamqptotalcertcns}{3131}
\DefineRemark{56715671ipv4tlsamqptotalcerts}{4193}
\DefineRemark{56715671ipv4tlsamqptotalconnerr}{5129014}
\DefineRemark{56715671ipv4tlsamqptotalconnsuccess}{216882}
\DefineRemark{56715671ipv4tlsamqptotalhosts}{5345896}
\DefineRemark{56715671ipv4tlsamqptotalips}{3696506441}
\DefineRemark{56715671ipv4tlsamqptotalsuccess}{28359}
\DefineRemark{56715671ipv4tlsamqptotalsystemid}{9147}
\DefineRemark{56715671ipv4tlsamqptotaltls}{205009}
\DefineRemark{56715671ipv4tlsamqptotaltlsauth}{204870}
\DefineRemark{56715671ipv4tlsamqptotaltlssuccess}{162492}
\DefineRemark{56715671ipv6amqpdate}{06-04}
\DefineRemark{56715671ipv6amqppctconnsuccess}{1.8}
\DefineRemark{56715671ipv6amqppcthostvalid}{0.2}
\DefineRemark{56715671ipv6amqppcttlsvalidoftransport}{0}
\DefineRemark{56715671ipv6amqppctvalidtls}{21}
\DefineRemark{56715671ipv6amqptotalas}{0}
\DefineRemark{56715671ipv6amqptotalasv6only}{2}
\DefineRemark{56715671ipv6amqptotalcertcns}{0}
\DefineRemark{56715671ipv6amqptotalcerts}{0}
\DefineRemark{56715671ipv6amqptotalconnerr}{2715896}
\DefineRemark{56715671ipv6amqptotalconnsuccess}{50538}
\DefineRemark{56715671ipv6amqptotalhosts}{2766434}
\DefineRemark{56715671ipv6amqptotalips}{1799131401}
\DefineRemark{56715671ipv6amqptotalsuccess}{0}
\DefineRemark{56715671ipv6amqptotalsystemid}{0}
\DefineRemark{56715671ipv6amqptotaltls}{0}
\DefineRemark{56715671ipv6amqptotaltlsauth}{0}
\DefineRemark{56715671ipv6amqptotaltlssuccess}{0}
\DefineRemark{56715671ipv6tlsamqpdate}{06-04}
\DefineRemark{56715671ipv6tlsamqppctconnsuccess}{1.8}
\DefineRemark{56715671ipv6tlsamqppcthostvalid}{0.2}
\DefineRemark{56715671ipv6tlsamqppcttlsvalidoftransport}{99}
\DefineRemark{56715671ipv6tlsamqppctvalidtls}{21}
\DefineRemark{56715671ipv6tlsamqptotalas}{43}
\DefineRemark{56715671ipv6tlsamqptotalasv6only}{2}
\DefineRemark{56715671ipv6tlsamqptotalcertcns}{112}
\DefineRemark{56715671ipv6tlsamqptotalcerts}{115}
\DefineRemark{56715671ipv6tlsamqptotalconnerr}{2715896}
\DefineRemark{56715671ipv6tlsamqptotalconnsuccess}{50538}
\DefineRemark{56715671ipv6tlsamqptotalhosts}{2766434}
\DefineRemark{56715671ipv6tlsamqptotalips}{1799131401}
\DefineRemark{56715671ipv6tlsamqptotalsuccess}{137}
\DefineRemark{56715671ipv6tlsamqptotalsystemid}{100}
\DefineRemark{56715671ipv6tlsamqptotaltls}{50236}
\DefineRemark{56715671ipv6tlsamqptotaltlsauth}{50236}
\DefineRemark{56715671ipv6tlsamqptotaltlssuccess}{41327}
\DefineRemark{56725672ipv4amqpdate}{06-04}
\DefineRemark{56725672ipv4amqppctconnsuccess}{5.9}
\DefineRemark{56725672ipv4amqppcthostvalid}{0.1}
\DefineRemark{56725672ipv4amqppcttlsvalidoftransport}{0}
\DefineRemark{56725672ipv4amqppctvalidtls}{21}
\DefineRemark{56725672ipv4amqptotalas}{5004}
\DefineRemark{56725672ipv4amqptotalasv6only}{0}
\DefineRemark{56725672ipv4amqptotalcertcns}{0}
\DefineRemark{56725672ipv4amqptotalcerts}{0}
\DefineRemark{56725672ipv4amqptotalconnerr}{4987662}
\DefineRemark{56725672ipv4amqptotalconnsuccess}{312883}
\DefineRemark{56725672ipv4amqptotalhosts}{5300545}
\DefineRemark{56725672ipv4amqptotalips}{3696506441}
\DefineRemark{56725672ipv4amqptotalsuccess}{143200}
\DefineRemark{56725672ipv4amqptotalsystemid}{92328}
\DefineRemark{56725672ipv4amqptotaltls}{0}
\DefineRemark{56725672ipv4amqptotaltlsauth}{0}
\DefineRemark{56725672ipv4amqptotaltlssuccess}{0}
\DefineRemark{56725672ipv4tlsamqpdate}{06-04}
\DefineRemark{56725672ipv4tlsamqppctconnsuccess}{5.9}
\DefineRemark{56725672ipv4tlsamqppcthostvalid}{0.1}
\DefineRemark{56725672ipv4tlsamqppcttlsvalidoftransport}{53}
\DefineRemark{56725672ipv4tlsamqppctvalidtls}{21}
\DefineRemark{56725672ipv4tlsamqptotalas}{538}
\DefineRemark{56725672ipv4tlsamqptotalasv6only}{0}
\DefineRemark{56725672ipv4tlsamqptotalcertcns}{660}
\DefineRemark{56725672ipv4tlsamqptotalcerts}{1133}
\DefineRemark{56725672ipv4tlsamqptotalconnerr}{4987662}
\DefineRemark{56725672ipv4tlsamqptotalconnsuccess}{312883}
\DefineRemark{56725672ipv4tlsamqptotalhosts}{5300545}
\DefineRemark{56725672ipv4tlsamqptotalips}{3696506441}
\DefineRemark{56725672ipv4tlsamqptotalsuccess}{11530}
\DefineRemark{56725672ipv4tlsamqptotalsystemid}{322}
\DefineRemark{56725672ipv4tlsamqptotaltls}{167107}
\DefineRemark{56725672ipv4tlsamqptotaltlsauth}{166736}
\DefineRemark{56725672ipv4tlsamqptotaltlssuccess}{128476}
\DefineRemark{56725672ipv6amqpdate}{06-06}
\DefineRemark{56725672ipv6amqppctconnsuccess}{6.4}
\DefineRemark{56725672ipv6amqppcthostvalid}{0}
\DefineRemark{56725672ipv6amqppcttlsvalidoftransport}{0}
\DefineRemark{56725672ipv6amqppctvalidtls}{21}
\DefineRemark{56725672ipv6amqptotalas}{305}
\DefineRemark{56725672ipv6amqptotalasv6only}{34}
\DefineRemark{56725672ipv6amqptotalcertcns}{0}
\DefineRemark{56725672ipv6amqptotalcerts}{0}
\DefineRemark{56725672ipv6amqptotalconnerr}{761113}
\DefineRemark{56725672ipv6amqptotalconnsuccess}{52150}
\DefineRemark{56725672ipv6amqptotalhosts}{813263}
\DefineRemark{56725672ipv6amqptotalips}{1849337203}
\DefineRemark{56725672ipv6amqptotalsuccess}{4664}
\DefineRemark{56725672ipv6amqptotalsystemid}{3764}
\DefineRemark{56725672ipv6amqptotaltls}{0}
\DefineRemark{56725672ipv6amqptotaltlsauth}{0}
\DefineRemark{56725672ipv6amqptotaltlssuccess}{0}
\DefineRemark{56725672ipv6tlsamqpdate}{06-06}
\DefineRemark{56725672ipv6tlsamqppctconnsuccess}{6.4}
\DefineRemark{56725672ipv6tlsamqppcthostvalid}{0}
\DefineRemark{56725672ipv6tlsamqppcttlsvalidoftransport}{90}
\DefineRemark{56725672ipv6tlsamqppctvalidtls}{21}
\DefineRemark{56725672ipv6tlsamqptotalas}{9}
\DefineRemark{56725672ipv6tlsamqptotalasv6only}{1}
\DefineRemark{56725672ipv6tlsamqptotalcertcns}{18}
\DefineRemark{56725672ipv6tlsamqptotalcerts}{18}
\DefineRemark{56725672ipv6tlsamqptotalconnerr}{761113}
\DefineRemark{56725672ipv6tlsamqptotalconnsuccess}{52150}
\DefineRemark{56725672ipv6tlsamqptotalhosts}{813263}
\DefineRemark{56725672ipv6tlsamqptotalips}{1849337203}
\DefineRemark{56725672ipv6tlsamqptotalsuccess}{18}
\DefineRemark{56725672ipv6tlsamqptotalsystemid}{13}
\DefineRemark{56725672ipv6tlsamqptotaltls}{47095}
\DefineRemark{56725672ipv6tlsamqptotaltlsauth}{47081}
\DefineRemark{56725672ipv6tlsamqptotaltlssuccess}{38217}
\DefineRemark{56835683ipv4coapdate}{06-18}
\DefineRemark{56835683ipv4coappctconnsuccess}{83}
\DefineRemark{56835683ipv4coappcthostvalid}{0}
\DefineRemark{56835683ipv4coappcttlsvalidoftransport}{0}
\DefineRemark{56835683ipv4coappctvalidtls}{0}
\DefineRemark{56835683ipv4coaptotalas}{3065}
\DefineRemark{56835683ipv4coaptotalasv6only}{0}
\DefineRemark{56835683ipv4coaptotalcertcns}{0}
\DefineRemark{56835683ipv4coaptotalcerts}{0}
\DefineRemark{56835683ipv4coaptotalconnerr}{57105}
\DefineRemark{56835683ipv4coaptotalconnsuccess}{269589}
\DefineRemark{56835683ipv4coaptotalhosts}{326694}
\DefineRemark{56835683ipv4coaptotalips}{3696506441}
\DefineRemark{56835683ipv4coaptotalsuccess}{266037}
\DefineRemark{56835683ipv4coaptotalsystemid}{336}
\DefineRemark{56835683ipv4coaptotaltls}{0}
\DefineRemark{56835683ipv4coaptotaltlsauth}{0}
\DefineRemark{56835683ipv4coaptotaltlssuccess}{0}
\DefineRemark{56835683ipv4tlscoapdate}{06-18}
\DefineRemark{56835683ipv4tlscoappctconnsuccess}{83}
\DefineRemark{56835683ipv4tlscoappcthostvalid}{0}
\DefineRemark{56835683ipv4tlscoappcttlsvalidoftransport}{0}
\DefineRemark{56835683ipv4tlscoappctvalidtls}{0}
\DefineRemark{56835683ipv4tlscoaptotalas}{0}
\DefineRemark{56835683ipv4tlscoaptotalasv6only}{0}
\DefineRemark{56835683ipv4tlscoaptotalcertcns}{0}
\DefineRemark{56835683ipv4tlscoaptotalcerts}{0}
\DefineRemark{56835683ipv4tlscoaptotalconnerr}{57105}
\DefineRemark{56835683ipv4tlscoaptotalconnsuccess}{269589}
\DefineRemark{56835683ipv4tlscoaptotalhosts}{326694}
\DefineRemark{56835683ipv4tlscoaptotalips}{3696506441}
\DefineRemark{56835683ipv4tlscoaptotalsuccess}{0}
\DefineRemark{56835683ipv4tlscoaptotalsystemid}{0}
\DefineRemark{56835683ipv4tlscoaptotaltls}{3}
\DefineRemark{56835683ipv4tlscoaptotaltlsauth}{3}
\DefineRemark{56835683ipv4tlscoaptotaltlssuccess}{0}
\DefineRemark{56845684ipv4coapdate}{06-17}
\DefineRemark{56845684ipv4coappctconnsuccess}{27}
\DefineRemark{56845684ipv4coappcthostvalid}{0}
\DefineRemark{56845684ipv4coappcttlsvalidoftransport}{0}
\DefineRemark{56845684ipv4coappctvalidtls}{0}
\DefineRemark{56845684ipv4coaptotalas}{0}
\DefineRemark{56845684ipv4coaptotalasv6only}{0}
\DefineRemark{56845684ipv4coaptotalcertcns}{0}
\DefineRemark{56845684ipv4coaptotalcerts}{0}
\DefineRemark{56845684ipv4coaptotalconnerr}{52952}
\DefineRemark{56845684ipv4coaptotalconnsuccess}{19889}
\DefineRemark{56845684ipv4coaptotalhosts}{72841}
\DefineRemark{56845684ipv4coaptotalips}{3696506441}
\DefineRemark{56845684ipv4coaptotalsuccess}{0}
\DefineRemark{56845684ipv4coaptotalsystemid}{0}
\DefineRemark{56845684ipv4coaptotaltls}{0}
\DefineRemark{56845684ipv4coaptotaltlsauth}{0}
\DefineRemark{56845684ipv4coaptotaltlssuccess}{0}
\DefineRemark{56845684ipv4tlscoapdate}{06-17}
\DefineRemark{56845684ipv4tlscoappctconnsuccess}{27}
\DefineRemark{56845684ipv4tlscoappcthostvalid}{0}
\DefineRemark{56845684ipv4tlscoappcttlsvalidoftransport}{3.5}
\DefineRemark{56845684ipv4tlscoappctvalidtls}{0}
\DefineRemark{56845684ipv4tlscoaptotalas}{23}
\DefineRemark{56845684ipv4tlscoaptotalasv6only}{0}
\DefineRemark{56845684ipv4tlscoaptotalcertcns}{43}
\DefineRemark{56845684ipv4tlscoaptotalcerts}{48}
\DefineRemark{56845684ipv4tlscoaptotalconnerr}{52952}
\DefineRemark{56845684ipv4tlscoaptotalconnsuccess}{19889}
\DefineRemark{56845684ipv4tlscoaptotalhosts}{72841}
\DefineRemark{56845684ipv4tlscoaptotalips}{3696506441}
\DefineRemark{56845684ipv4tlscoaptotalsuccess}{91}
\DefineRemark{56845684ipv4tlscoaptotalsystemid}{28}
\DefineRemark{56845684ipv4tlscoaptotaltls}{702}
\DefineRemark{56845684ipv4tlscoaptotaltlsauth}{702}
\DefineRemark{56845684ipv4tlscoaptotaltlssuccess}{600}
\DefineRemark{56845684ipv6coapdate}{06-23}
\DefineRemark{56845684ipv6coappctconnsuccess}{44}
\DefineRemark{56845684ipv6coappcthostvalid}{0}
\DefineRemark{56845684ipv6coappcttlsvalidoftransport}{0}
\DefineRemark{56845684ipv6coappctvalidtls}{0}
\DefineRemark{56845684ipv6coaptotalas}{0}
\DefineRemark{56845684ipv6coaptotalasv6only}{0}
\DefineRemark{56845684ipv6coaptotalcertcns}{0}
\DefineRemark{56845684ipv6coaptotalcerts}{0}
\DefineRemark{56845684ipv6coaptotalconnerr}{31}
\DefineRemark{56845684ipv6coaptotalconnsuccess}{24}
\DefineRemark{56845684ipv6coaptotalhosts}{55}
\DefineRemark{56845684ipv6coaptotalips}{1762033586}
\DefineRemark{56845684ipv6coaptotalsuccess}{0}
\DefineRemark{56845684ipv6coaptotalsystemid}{0}
\DefineRemark{56845684ipv6coaptotaltls}{0}
\DefineRemark{56845684ipv6coaptotaltlsauth}{0}
\DefineRemark{56845684ipv6coaptotaltlssuccess}{0}
\DefineRemark{56845684ipv6tlscoapdate}{06-23}
\DefineRemark{56845684ipv6tlscoappctconnsuccess}{44}
\DefineRemark{56845684ipv6tlscoappcthostvalid}{0}
\DefineRemark{56845684ipv6tlscoappcttlsvalidoftransport}{42}
\DefineRemark{56845684ipv6tlscoappctvalidtls}{0}
\DefineRemark{56845684ipv6tlscoaptotalas}{2}
\DefineRemark{56845684ipv6tlscoaptotalasv6only}{0}
\DefineRemark{56845684ipv6tlscoaptotalcertcns}{2}
\DefineRemark{56845684ipv6tlscoaptotalcerts}{2}
\DefineRemark{56845684ipv6tlscoaptotalconnerr}{31}
\DefineRemark{56845684ipv6tlscoaptotalconnsuccess}{24}
\DefineRemark{56845684ipv6tlscoaptotalhosts}{55}
\DefineRemark{56845684ipv6tlscoaptotalips}{1762033586}
\DefineRemark{56845684ipv6tlscoaptotalsuccess}{2}
\DefineRemark{56845684ipv6tlscoaptotalsystemid}{2}
\DefineRemark{56845684ipv6tlscoaptotaltls}{10}
\DefineRemark{56845684ipv6tlscoaptotaltlsauth}{10}
\DefineRemark{56845684ipv6tlscoaptotaltlssuccess}{10}
\DefineRemark{88838883ipv4mqttdate}{05-27}
\DefineRemark{88838883ipv4mqttpctconnsuccess}{12}
\DefineRemark{88838883ipv4mqttpcthostvalid}{0.2}
\DefineRemark{88838883ipv4mqttpcttlsvalidoftransport}{0}
\DefineRemark{88838883ipv4mqttpctvalidtls}{3.6}
\DefineRemark{88838883ipv4mqtttotalas}{0}
\DefineRemark{88838883ipv4mqtttotalasv6only}{0}
\DefineRemark{88838883ipv4mqtttotalcertcns}{0}
\DefineRemark{88838883ipv4mqtttotalcerts}{0}
\DefineRemark{88838883ipv4mqtttotalconnerr}{5299438}
\DefineRemark{88838883ipv4mqtttotalconnsuccess}{704098}
\DefineRemark{88838883ipv4mqtttotalhosts}{6003536}
\DefineRemark{88838883ipv4mqtttotalips}{3696506441}
\DefineRemark{88838883ipv4mqtttotalsuccess}{0}
\DefineRemark{88838883ipv4mqtttotalsystemid}{0}
\DefineRemark{88838883ipv4mqtttotaltls}{0}
\DefineRemark{88838883ipv4mqtttotaltlsauth}{0}
\DefineRemark{88838883ipv4mqtttotaltlssuccess}{0}
\DefineRemark{88838883ipv4tlsmqttdate}{05-27}
\DefineRemark{88838883ipv4tlsmqttpctconnsuccess}{12}
\DefineRemark{88838883ipv4tlsmqttpcthostvalid}{0.2}
\DefineRemark{88838883ipv4tlsmqttpcttlsvalidoftransport}{34}
\DefineRemark{88838883ipv4tlsmqttpctvalidtls}{3.6}
\DefineRemark{88838883ipv4tlsmqtttotalas}{881}
\DefineRemark{88838883ipv4tlsmqtttotalasv6only}{0}
\DefineRemark{88838883ipv4tlsmqtttotalcertcns}{3690}
\DefineRemark{88838883ipv4tlsmqtttotalcerts}{3983}
\DefineRemark{88838883ipv4tlsmqtttotalconnerr}{5299438}
\DefineRemark{88838883ipv4tlsmqtttotalconnsuccess}{704098}
\DefineRemark{88838883ipv4tlsmqtttotalhosts}{6003536}
\DefineRemark{88838883ipv4tlsmqtttotalips}{3696506441}
\DefineRemark{88838883ipv4tlsmqtttotalsuccess}{12759}
\DefineRemark{88838883ipv4tlsmqtttotalsystemid}{0}
\DefineRemark{88838883ipv4tlsmqtttotaltls}{236328}
\DefineRemark{88838883ipv4tlsmqtttotaltlsauth}{236328}
\DefineRemark{88838883ipv4tlsmqtttotaltlssuccess}{232014}
\DefineRemark{88838883ipv6mqttdate}{05-29}
\DefineRemark{88838883ipv6mqttpctconnsuccess}{18}
\DefineRemark{88838883ipv6mqttpcthostvalid}{0.1}
\DefineRemark{88838883ipv6mqttpcttlsvalidoftransport}{0}
\DefineRemark{88838883ipv6mqttpctvalidtls}{3.6}
\DefineRemark{88838883ipv6mqtttotalas}{0}
\DefineRemark{88838883ipv6mqtttotalasv6only}{17}
\DefineRemark{88838883ipv6mqtttotalcertcns}{0}
\DefineRemark{88838883ipv6mqtttotalcerts}{0}
\DefineRemark{88838883ipv6mqtttotalconnerr}{876177}
\DefineRemark{88838883ipv6mqtttotalconnsuccess}{186269}
\DefineRemark{88838883ipv6mqtttotalhosts}{1062446}
\DefineRemark{88838883ipv6mqtttotalips}{1800732625}
\DefineRemark{88838883ipv6mqtttotalsuccess}{0}
\DefineRemark{88838883ipv6mqtttotalsystemid}{0}
\DefineRemark{88838883ipv6mqtttotaltls}{0}
\DefineRemark{88838883ipv6mqtttotaltlsauth}{0}
\DefineRemark{88838883ipv6mqtttotaltlssuccess}{0}
\DefineRemark{88838883ipv6tlsmqttdate}{05-29}
\DefineRemark{88838883ipv6tlsmqttpctconnsuccess}{18}
\DefineRemark{88838883ipv6tlsmqttpcthostvalid}{0.1}
\DefineRemark{88838883ipv6tlsmqttpcttlsvalidoftransport}{30}
\DefineRemark{88838883ipv6tlsmqttpctvalidtls}{3.6}
\DefineRemark{88838883ipv6tlsmqtttotalas}{68}
\DefineRemark{88838883ipv6tlsmqtttotalasv6only}{15}
\DefineRemark{88838883ipv6tlsmqtttotalcertcns}{190}
\DefineRemark{88838883ipv6tlsmqtttotalcerts}{193}
\DefineRemark{88838883ipv6tlsmqtttotalconnerr}{876177}
\DefineRemark{88838883ipv6tlsmqtttotalconnsuccess}{186269}
\DefineRemark{88838883ipv6tlsmqtttotalhosts}{1062446}
\DefineRemark{88838883ipv6tlsmqtttotalips}{1800732625}
\DefineRemark{88838883ipv6tlsmqtttotalsuccess}{245}
\DefineRemark{88838883ipv6tlsmqtttotalsystemid}{0}
\DefineRemark{88838883ipv6tlsmqtttotaltls}{55179}
\DefineRemark{88838883ipv6tlsmqtttotaltlsauth}{55179}
\DefineRemark{88838883ipv6tlsmqtttotaltlssuccess}{55178}
\DefineRemark{allamqppctvalidtls}{21} %
\DefineRemark{allcoappctvalidtls}{0} %
\DefineRemark{allmqttpctvalidtls}{3.6} %
\DefineRemark{allopcuapctvalidtls}{0.3} %
\DefineRemark{alltlsamqptotalas}{1102} %
\DefineRemark{alltlsamqptotalcertcns}{817} %
\DefineRemark{alltlsamqptotalcertcnsother}{3} %
\DefineRemark{alltlsamqptotalsuccess}{37656} %
\DefineRemark{alltlscoaptotalas}{23} %
\DefineRemark{alltlscoaptotalcertcns}{45} %
\DefineRemark{alltlscoaptotalcertcnsother}{0} %
\DefineRemark{alltlscoaptotalsuccess}{93} %
\DefineRemark{alltlsmqtttotalas}{903} %
\DefineRemark{alltlsmqtttotalcertcns}{2058} %
\DefineRemark{alltlsmqtttotalcertcnsother}{12} %
\DefineRemark{alltlsmqtttotalsuccess}{13401} %
\DefineRemark{alltlsopcuatotalas}{1} %
\DefineRemark{alltlsopcuatotalcertcns}{5} %
\DefineRemark{alltlsopcuatotalcertcnsother}{864} %
\DefineRemark{alltlsopcuatotalsuccess}{5} %
\DefineRemark{backendipv4pctvalidtls}{9.4} %
\DefineRemark{backendipv4totalsuccess}{539719} %
\DefineRemark{backendipv6pctvalidtls}{6.2} %
\DefineRemark{backendipv6totalsuccess}{6650} %
\DefineRemark{backendtlsipv4totalas}{7298} %
\DefineRemark{backendtlsipv4totalastls}{1696} %
\DefineRemark{backendtlsipv4totalcertcns}{7367} %
\DefineRemark{backendtlsipv4totalsuccess}{50616} %
\DefineRemark{backendtlsipv6totalas}{426} %
\DefineRemark{backendtlsipv6totalastls}{101} %
\DefineRemark{backendtlsipv6totalcertcns}{321} %
\DefineRemark{backendtlsipv6totalsuccess}{410} %
\DefineRemark{deviceipv4pctvalidtls}{0} %
\DefineRemark{deviceipv4totalsuccess}{267865} %
\DefineRemark{deviceipv6pctvalidtls}{25} %
\DefineRemark{deviceipv6totalsuccess}{8} %
\DefineRemark{devicetlsipv4totalas}{3317} %
\DefineRemark{devicetlsipv4totalastls}{23} %
\DefineRemark{devicetlsipv4totalcertcns}{48} %
\DefineRemark{devicetlsipv4totalsuccess}{96} %
\DefineRemark{devicetlsipv6totalas}{5} %
\DefineRemark{devicetlsipv6totalastls}{2} %
\DefineRemark{devicetlsipv6totalcertcns}{2} %
\DefineRemark{devicetlsipv6totalsuccess}{2} %
\DefineRemark{ipv4amqpasperhost}{0.03} %
\DefineRemark{ipv4amqpastyperank0}{Content} %
\DefineRemark{ipv4amqpastyperank1}{Enterprise} %
\DefineRemark{ipv4amqpastyperank2}{NSP} %
\DefineRemark{ipv4amqpastyperank3}{ISP} %
\DefineRemark{ipv4amqpastyperank4}{Education} %
\DefineRemark{ipv4amqpastyperank5}{Network Services} %
\DefineRemark{ipv4amqpastyperank6}{Non-Profit} %
\DefineRemark{ipv4amqpastyperank7}{Route Server} %
\DefineRemark{ipv4amqpastyperank8}{Government} %
\DefineRemark{ipv4amqpcnperhost}{0.02} %
\DefineRemark{ipv4amqppctaliased}{0} %
\DefineRemark{ipv4amqppctastyperank0}{74} %
\DefineRemark{ipv4amqppctastyperank1}{15} %
\DefineRemark{ipv4amqppctastyperank2}{6.4} %
\DefineRemark{ipv4amqppctastyperank3}{3.7} %
\DefineRemark{ipv4amqppctastyperank4}{0.5} %
\DefineRemark{ipv4amqppctastyperank5}{0.2} %
\DefineRemark{ipv4amqppctastyperank6}{0} %
\DefineRemark{ipv4amqppctastyperank7}{0} %
\DefineRemark{ipv4amqppctastyperank8}{0} %
\DefineRemark{ipv4amqppctvalidtls}{22} %
\DefineRemark{ipv4amqptotalaliased}{0} %
\DefineRemark{ipv4amqptotalas}{5362} %
\DefineRemark{ipv4amqptotalasv6only}{0} %
\DefineRemark{ipv4amqptotalcertcns}{3465} %
\DefineRemark{ipv4coapasperhost}{0.01} %
\DefineRemark{ipv4coapastyperank0}{NSP} %
\DefineRemark{ipv4coapastyperank1}{ISP} %
\DefineRemark{ipv4coapastyperank2}{Content} %
\DefineRemark{ipv4coapastyperank3}{Enterprise} %
\DefineRemark{ipv4coapastyperank4}{Education} %
\DefineRemark{ipv4coapastyperank5}{Network Services} %
\DefineRemark{ipv4coapastyperank6}{Non-Profit} %
\DefineRemark{ipv4coapastyperank7}{Route Server} %
\DefineRemark{ipv4coapcnperhost}{0} %
\DefineRemark{ipv4coappctaliased}{0} %
\DefineRemark{ipv4coappctastyperank0}{66} %
\DefineRemark{ipv4coappctastyperank1}{32} %
\DefineRemark{ipv4coappctastyperank2}{1.3} %
\DefineRemark{ipv4coappctastyperank3}{0.6} %
\DefineRemark{ipv4coappctastyperank4}{0} %
\DefineRemark{ipv4coappctastyperank5}{0} %
\DefineRemark{ipv4coappctastyperank6}{0} %
\DefineRemark{ipv4coappctastyperank7}{0} %
\DefineRemark{ipv4coappctvalidtls}{0} %
\DefineRemark{ipv4coaptotalaliased}{0} %
\DefineRemark{ipv4coaptotalas}{3067} %
\DefineRemark{ipv4coaptotalasv6only}{0} %
\DefineRemark{ipv4coaptotalcertcns}{43} %
\DefineRemark{ipv4mqttasperhost}{0.01} %
\DefineRemark{ipv4mqttastyperank0}{ISP} %
\DefineRemark{ipv4mqttastyperank1}{Content} %
\DefineRemark{ipv4mqttastyperank2}{NSP} %
\DefineRemark{ipv4mqttastyperank3}{Enterprise} %
\DefineRemark{ipv4mqttastyperank4}{Education} %
\DefineRemark{ipv4mqttastyperank5}{Network Services} %
\DefineRemark{ipv4mqttastyperank6}{Non-Profit} %
\DefineRemark{ipv4mqttastyperank7}{Government} %
\DefineRemark{ipv4mqttcnperhost}{0.01} %
\DefineRemark{ipv4mqttpctaliased}{0} %
\DefineRemark{ipv4mqttpctastyperank0}{80} %
\DefineRemark{ipv4mqttpctastyperank1}{10} %
\DefineRemark{ipv4mqttpctastyperank2}{7.4} %
\DefineRemark{ipv4mqttpctastyperank3}{2.2} %
\DefineRemark{ipv4mqttpctastyperank4}{0.2} %
\DefineRemark{ipv4mqttpctastyperank5}{0} %
\DefineRemark{ipv4mqttpctastyperank6}{0} %
\DefineRemark{ipv4mqttpctastyperank7}{0} %
\DefineRemark{ipv4mqttpctvalidtls}{3.5} %
\DefineRemark{ipv4mqtttotalaliased}{0} %
\DefineRemark{ipv4mqtttotalas}{4060} %
\DefineRemark{ipv4mqtttotalasv6only}{0} %
\DefineRemark{ipv4mqtttotalcertcns}{3938} %
\DefineRemark{ipv4opcuaasperhost}{0.28} %
\DefineRemark{ipv4opcuaastyperank0}{ISP} %
\DefineRemark{ipv4opcuaastyperank1}{NSP} %
\DefineRemark{ipv4opcuaastyperank2}{Content} %
\DefineRemark{ipv4opcuaastyperank3}{Education} %
\DefineRemark{ipv4opcuaastyperank4}{Enterprise} %
\DefineRemark{ipv4opcuaastyperank5}{Network Services} %
\DefineRemark{ipv4opcuaastyperank6}{Non-Profit} %
\DefineRemark{ipv4opcuacnperhost}{0.5} %
\DefineRemark{ipv4opcuapctaliased}{0} %
\DefineRemark{ipv4opcuapctastyperank0}{38} %
\DefineRemark{ipv4opcuapctastyperank1}{35} %
\DefineRemark{ipv4opcuapctastyperank2}{15} %
\DefineRemark{ipv4opcuapctastyperank3}{6} %
\DefineRemark{ipv4opcuapctastyperank4}{5.2} %
\DefineRemark{ipv4opcuapctastyperank5}{0.5} %
\DefineRemark{ipv4opcuapctastyperank6}{0.2} %
\DefineRemark{ipv4opcuapctvalidtls}{0.3} %
\DefineRemark{ipv4opcuatotalaliased}{0} %
\DefineRemark{ipv4opcuatotalas}{491} %
\DefineRemark{ipv4opcuatotalasv6only}{0} %
\DefineRemark{ipv4opcuatotalcertcns}{869} %
\DefineRemark{ipv4pctaliased}{0} %
\DefineRemark{ipv4pctscansuccess}{0}
\DefineRemark{ipv4pctvalidtls}{6.3} %
\DefineRemark{ipv4permillescansuccess}{0.013}
\DefineRemark{ipv4totalaliased}{0} %
\DefineRemark{ipv4totaldefaultsuccess}{766481} %
\DefineRemark{ipv4totalsuccess}{807230} %
\DefineRemark{ipv6amqpasperhost}{0.07} %
\DefineRemark{ipv6amqpastyperank0}{Content} %
\DefineRemark{ipv6amqpastyperank1}{NSP} %
\DefineRemark{ipv6amqpastyperank2}{ISP} %
\DefineRemark{ipv6amqpastyperank3}{Education} %
\DefineRemark{ipv6amqpastyperank4}{Enterprise} %
\DefineRemark{ipv6amqpastyperank5}{Non-Profit} %
\DefineRemark{ipv6amqpastyperank6}{Network Services} %
\DefineRemark{ipv6amqpcnperhost}{0.03} %
\DefineRemark{ipv6amqppctaliased}{0.5} %
\DefineRemark{ipv6amqppctastyperank0}{89} %
\DefineRemark{ipv6amqppctastyperank1}{5.9} %
\DefineRemark{ipv6amqppctastyperank2}{2.1} %
\DefineRemark{ipv6amqppctastyperank3}{1.4} %
\DefineRemark{ipv6amqppctastyperank4}{0.8} %
\DefineRemark{ipv6amqppctastyperank5}{0.7} %
\DefineRemark{ipv6amqppctastyperank6}{0.2} %
\DefineRemark{ipv6amqppctvalidtls}{3.2} %
\DefineRemark{ipv6amqptotalaliased}{25} %
\DefineRemark{ipv6amqptotalas}{318} %
\DefineRemark{ipv6amqptotalasv6only}{33} %
\DefineRemark{ipv6amqptotalcertcns}{126} %
\DefineRemark{ipv6coapasperhost}{1} %
\DefineRemark{ipv6coapastyperank0}{Content} %
\DefineRemark{ipv6coapastyperank1}{Education} %
\DefineRemark{ipv6coapastyperank2}{N/A} %
\DefineRemark{ipv6coapcnperhost}{1} %
\DefineRemark{ipv6coappctaliased}{0} %
\DefineRemark{ipv6coappctastyperank0}{50} %
\DefineRemark{ipv6coappctastyperank1}{50} %
\DefineRemark{ipv6coappctastyperank2}{0} %
\DefineRemark{ipv6coappctvalidtls}{100} %
\DefineRemark{ipv6coaptotalaliased}{0} %
\DefineRemark{ipv6coaptotalas}{2} %
\DefineRemark{ipv6coaptotalasv6only}{0} %
\DefineRemark{ipv6coaptotalcertcns}{2} %
\DefineRemark{ipv6mqttasperhost}{0.11} %
\DefineRemark{ipv6mqttastyperank0}{Content} %
\DefineRemark{ipv6mqttastyperank1}{NSP} %
\DefineRemark{ipv6mqttastyperank2}{ISP} %
\DefineRemark{ipv6mqttastyperank3}{Enterprise} %
\DefineRemark{ipv6mqttastyperank4}{Education} %
\DefineRemark{ipv6mqttastyperank5}{Non-Profit} %
\DefineRemark{ipv6mqttastyperank6}{Network Services} %
\DefineRemark{ipv6mqttcnperhost}{0.1} %
\DefineRemark{ipv6mqttpctaliased}{1} %
\DefineRemark{ipv6mqttpctastyperank0}{75} %
\DefineRemark{ipv6mqttpctastyperank1}{12} %
\DefineRemark{ipv6mqttpctastyperank2}{8.1} %
\DefineRemark{ipv6mqttpctastyperank3}{1.9} %
\DefineRemark{ipv6mqttpctastyperank4}{1.8} %
\DefineRemark{ipv6mqttpctastyperank5}{0.8} %
\DefineRemark{ipv6mqttpctastyperank6}{0.1} %
\DefineRemark{ipv6mqttpctvalidtls}{13} %
\DefineRemark{ipv6mqtttotalaliased}{20} %
\DefineRemark{ipv6mqtttotalas}{221} %
\DefineRemark{ipv6mqtttotalasv6only}{8} %
\DefineRemark{ipv6mqtttotalcertcns}{197} %
\DefineRemark{ipv6opcuaasperhost}{0.5} %
\DefineRemark{ipv6opcuaastyperank0}{Content} %
\DefineRemark{ipv6opcuaastyperank1}{NSP} %
\DefineRemark{ipv6opcuaastyperank2}{N/A} %
\DefineRemark{ipv6opcuacnperhost}{0} %
\DefineRemark{ipv6opcuapctaliased}{0} %
\DefineRemark{ipv6opcuapctastyperank0}{83} %
\DefineRemark{ipv6opcuapctastyperank1}{17} %
\DefineRemark{ipv6opcuapctastyperank2}{0} %
\DefineRemark{ipv6opcuapctvalidtls}{0} %
\DefineRemark{ipv6opcuatotalaliased}{0} %
\DefineRemark{ipv6opcuatotalas}{3} %
\DefineRemark{ipv6opcuatotalasv6only}{0} %
\DefineRemark{ipv6opcuatotalcertcns}{0} %
\DefineRemark{ipv6pctaliased}{0.7} %
\DefineRemark{ipv6pctscansuccess}{0}
\DefineRemark{ipv6pctvalidtls}{6.2} %
\DefineRemark{ipv6permillescansuccess}{0.000248}
\DefineRemark{ipv6totalaliased}{45} %
\DefineRemark{ipv6totaldefaultsuccess}{6249} %
\DefineRemark{ipv6totalsuccess}{6658} %
\DefineRemark{optionaltlsipv4amqptotalsuccess}{19702} %
\DefineRemark{optionaltlsipv4coaptotalsuccess}{50} %
\DefineRemark{optionaltlsipv4mqtttotalsuccess}{9104} %
\DefineRemark{optionaltlsipv4opcuatotalsuccess}{0} %
\DefineRemark{optionaltlsipv6amqptotalsuccess}{41} %
\DefineRemark{optionaltlsipv6coaptotalsuccess}{0} %
\DefineRemark{optionaltlsipv6mqtttotalsuccess}{119} %
\DefineRemark{optionaltlsipv6opcuatotalsuccess}{0} %
\DefineRemark{scannedipv6addresses}{14894862320}
\DefineRemark{tlsipv4amqpcnperhost}{0.09} %
\DefineRemark{tlsipv4amqptotalastls}{1096} %
\DefineRemark{tlsipv4amqptotalsuccess}{37501} %
\DefineRemark{tlsipv4coapcnperhost}{0.47} %
\DefineRemark{tlsipv4coaptotalastls}{23} %
\DefineRemark{tlsipv4coaptotalsuccess}{91} %
\DefineRemark{tlsipv4mqttcnperhost}{0.3} %
\DefineRemark{tlsipv4mqtttotalastls}{897} %
\DefineRemark{tlsipv4mqtttotalsuccess}{13145} %
\DefineRemark{tlsipv4opcuacnperhost}{173.8} %
\DefineRemark{tlsipv4opcuatotalastls}{1} %
\DefineRemark{tlsipv4opcuatotalsuccess}{5} %
\DefineRemark{tlsipv4totalas}{9161} %
\DefineRemark{tlsipv4totalastls}{1696} %
\DefineRemark{tlsipv4totalcertcns}{7407} %
\DefineRemark{tlsipv4totalsuccess}{50712} %
\DefineRemark{tlsipv6amqpcnperhost}{0.81} %
\DefineRemark{tlsipv6amqptotalastls}{46} %
\DefineRemark{tlsipv6amqptotalsuccess}{155} %
\DefineRemark{tlsipv6coapcnperhost}{1} %
\DefineRemark{tlsipv6coaptotalastls}{2} %
\DefineRemark{tlsipv6coaptotalsuccess}{2} %
\DefineRemark{tlsipv6mqttcnperhost}{0.77} %
\DefineRemark{tlsipv6mqtttotalastls}{74} %
\DefineRemark{tlsipv6mqtttotalsuccess}{256} %
\DefineRemark{tlsipv6opcuacnperhost}{0} %
\DefineRemark{tlsipv6opcuatotalastls}{0} %
\DefineRemark{tlsipv6opcuatotalsuccess}{0} %
\DefineRemark{tlsipv6totalas}{426} %
\DefineRemark{tlsipv6totalastls}{102} %
\DefineRemark{tlsipv6totalcertcns}{322} %
\DefineRemark{tlsipv6totalsuccess}{412} %

%% file: variables/textvalues.tex
\DefineRemark{hostscnttlsweak}{19175} %
\DefineRemark{hostspcttlsweak}{38} %
\DefineRemark{hostspcttlsweakduetoauth}{26} %
\DefineRemark{hostspcttlsweakduetocipher}{1.3} %
\DefineRemark{hostspcttlsweakduetodeprecatedcert}{1} %
\DefineRemark{hostspcttlsweakduetoreuse}{28} %
\DefineRemark{hostspcttlsweakduetoversion}{0.1} %
\DefineRemark{ipv6deploymentsinipv6onlyases}{89}
\DefineRemark{ipv6hostscnttlsweak}{312} %
\DefineRemark{ipv6hostspcttlsweak}{76} %
\DefineRemark{ipv6hostspcttlsweakduetoauth}{65} %
\DefineRemark{ipv6hostspcttlsweakduetocipher}{5.3} %
\DefineRemark{ipv6hostspcttlsweakduetodeprecatedcert}{4.9} %
\DefineRemark{ipv6hostspcttlsweakduetoreuse}{20} %
\DefineRemark{ipv6hostspcttlsweakduetoversion}{1.2} %

%% file: variables/addressorigin.tex
\DefineRemark{generatorvaliddeploymentsDNS6Forest}{161}
\DefineRemark{generatorvaliddeploymentsDNS6GAN}{0}
\DefineRemark{generatorvaliddeploymentsDNS6GCVAE}{0}
\DefineRemark{generatorvaliddeploymentsDNS6Gen}{61}
\DefineRemark{generatorvaliddeploymentsDNS6Graph}{262}
\DefineRemark{generatorvaliddeploymentsDNS6Hit}{5}
\DefineRemark{generatorvaliddeploymentsDNS6Scan}{250}
\DefineRemark{generatorvaliddeploymentsDNS6Tree}{245}
\DefineRemark{generatorvaliddeploymentsDNS6VecLM}{0}
\DefineRemark{generatorvaliddeploymentsDNSEIPGen}{0}
\DefineRemark{generatorvaliddeploymentsDNSEIPGenerator}{0}
\DefineRemark{generatorvaliddeploymentsDNSEntropyIP}{0}
\DefineRemark{generatorvaliddeploymentsDNSwww6Forest}{41}
\DefineRemark{generatorvaliddeploymentsDNSwww6GAN}{0}
\DefineRemark{generatorvaliddeploymentsDNSwww6GCVAE}{0}
\DefineRemark{generatorvaliddeploymentsDNSwww6Gen}{27}
\DefineRemark{generatorvaliddeploymentsDNSwww6Graph}{296}
\DefineRemark{generatorvaliddeploymentsDNSwww6Hit}{0}
\DefineRemark{generatorvaliddeploymentsDNSwww6Scan}{282}
\DefineRemark{generatorvaliddeploymentsDNSwww6Tree}{261}
\DefineRemark{generatorvaliddeploymentsDNSwww6VecLM}{0}
\DefineRemark{generatorvaliddeploymentsDNSwwwEIPGen}{0}
\DefineRemark{generatorvaliddeploymentsDNSwwwEIPGenerator}{0}
\DefineRemark{generatorvaliddeploymentsDNSwwwEntropyIP}{0}
\DefineRemark{generatorvaliddeploymentsTUM6Forest}{0}
\DefineRemark{generatorvaliddeploymentsTUM6GAN}{0}
\DefineRemark{generatorvaliddeploymentsTUM6GCVAE}{0}
\DefineRemark{generatorvaliddeploymentsTUM6Gen}{12}
\DefineRemark{generatorvaliddeploymentsTUM6Graph}{13}
\DefineRemark{generatorvaliddeploymentsTUM6Hit}{0}
\DefineRemark{generatorvaliddeploymentsTUM6Scan}{0}
\DefineRemark{generatorvaliddeploymentsTUM6Tree}{56}
\DefineRemark{generatorvaliddeploymentsTUM6VecLM}{0}
\DefineRemark{generatorvaliddeploymentsTUMEIPGen}{0}
\DefineRemark{generatorvaliddeploymentsTUMEIPGenerator}{0}
\DefineRemark{generatorvaliddeploymentsTUMEntropyIP}{0}
\DefineRemark{generatorvaliddeploymentsTUMopen6Forest}{0}
\DefineRemark{generatorvaliddeploymentsTUMopen6GAN}{0}
\DefineRemark{generatorvaliddeploymentsTUMopen6GCVAE}{0}
\DefineRemark{generatorvaliddeploymentsTUMopen6Gen}{0}
\DefineRemark{generatorvaliddeploymentsTUMopen6Graph}{0}
\DefineRemark{generatorvaliddeploymentsTUMopen6Hit}{0}
\DefineRemark{generatorvaliddeploymentsTUMopen6Scan}{0}
\DefineRemark{generatorvaliddeploymentsTUMopen6Tree}{0}
\DefineRemark{generatorvaliddeploymentsTUMopen6VecLM}{0}
\DefineRemark{generatorvaliddeploymentsTUMopenEIPGen}{0}
\DefineRemark{generatorvaliddeploymentsTUMopenEIPGenerator}{0}
\DefineRemark{generatorvaliddeploymentsTUMopenEntropyIP}{0}
\DefineRemark{generatorvaliddeploymentsv46Forest}{120}
\DefineRemark{generatorvaliddeploymentsv46GAN}{9}
\DefineRemark{generatorvaliddeploymentsv46GCVAE}{7}
\DefineRemark{generatorvaliddeploymentsv46Gen}{58}
\DefineRemark{generatorvaliddeploymentsv46Graph}{45}
\DefineRemark{generatorvaliddeploymentsv46Hit}{0}
\DefineRemark{generatorvaliddeploymentsv46Scan}{66}
\DefineRemark{generatorvaliddeploymentsv46Tree}{77}
\DefineRemark{generatorvaliddeploymentsv46VecLM}{1}
\DefineRemark{generatorvaliddeploymentsv4EIPGen}{5}
\DefineRemark{generatorvaliddeploymentsv4EIPGenerator}{5}
\DefineRemark{generatorvaliddeploymentsv4EntropyIP}{7}
\DefineRemark{maxhitratepct}{0}
\DefineRemark{maxhitratepermille}{0.025}
\DefineRemark{originnumall}{6574}
\DefineRemark{originnumdnsAAAAwww6Scan6Tree}{6208}
\DefineRemark{originnumdnsAAAAwww6Scan6Tree6Graph}{6324}
\DefineRemark{originnumseedonly}{5890}
\DefineRemark{originpctnumdnsAAAAwww6Scan6Tree}{94}
\DefineRemark{originpctnumdnsAAAAwww6Scan6Tree6Graph}{96}
\DefineRemark{originpctnumseedonly}{90}
\DefineRemark{pcthostsSeedAnddnsAAAAwww6Graph}{95}
\DefineRemark{pcthostsSeedAnddnsAAAAwww6Scan}{93}
\DefineRemark{pcthostsSeedAnddnsAAAAwww6Scan6Graph}{95}
\DefineRemark{pcthostsSeedAnddnsAAAAwww6Scan6Tree6Graph}{95}
\DefineRemark{pcthostsSeedAnddnsAAAAwww6Tree}{94}
\DefineRemark{pcthostsdnsAAAAwww6Scan}{4.2}
\DefineRemark{pcthostsdnsAAAAwww6Scan6Graph}{6.1}
\DefineRemark{pcthostsdnsAAAAwww6Scan6Tree}{5.1}
\DefineRemark{pcthostsdnsAAAAwww6Scan6Tree6Graph}{6.1}
\DefineRemark{pcthostsgeneratorsonly}{12}
\DefineRemark{pcthostsseedsonly}{88}
\DefineRemark{seedlistgainDNSDNS}{366}
\DefineRemark{seedlistgainDNSDNSwww}{58}
\DefineRemark{seedlistgainDNSTUM}{5493}
\DefineRemark{seedlistgainDNSTUMopen}{4109}
\DefineRemark{seedlistgainDNSv4}{2056}
\DefineRemark{seedlistgainDNSwwwDNS}{77}
\DefineRemark{seedlistgainDNSwwwDNSwww}{347}
\DefineRemark{seedlistgainDNSwwwTUM}{5516}
\DefineRemark{seedlistgainDNSwwwTUMopen}{4131}
\DefineRemark{seedlistgainDNSwwwv4}{2064}
\DefineRemark{seedlistgainTUMDNS}{8}
\DefineRemark{seedlistgainTUMDNSwww}{12}
\DefineRemark{seedlistgainTUMTUM}{5849}
\DefineRemark{seedlistgainTUMTUMopen}{0}
\DefineRemark{seedlistgainTUMopenDNS}{68}
\DefineRemark{seedlistgainTUMopenDNSwww}{71}
\DefineRemark{seedlistgainTUMopenTUM}{1444}
\DefineRemark{seedlistgainTUMopenTUMopen}{4405}
\DefineRemark{seedlistgainTUMopenv4}{527}
\DefineRemark{seedlistgainTUMv4}{24}
\DefineRemark{seedlistgainv4DNS}{177}
\DefineRemark{seedlistgainv4DNSwww}{166}
\DefineRemark{seedlistgainv4TUM}{3629}
\DefineRemark{seedlistgainv4TUMopen}{2688}
\DefineRemark{seedlistgainv4v4}{2245}
\DefineRemark{totalhostsSeedAnddnsAAAAwww6Graph}{6294}
\DefineRemark{totalhostsSeedAnddnsAAAAwww6Scan}{6172}
\DefineRemark{totalhostsSeedAnddnsAAAAwww6Scan6Graph}{6294}
\DefineRemark{totalhostsSeedAnddnsAAAAwww6Scan6Tree6Graph}{6338}
\DefineRemark{totalhostsSeedAnddnsAAAAwww6Tree}{6227}
\DefineRemark{totalhostsdnsAAAAwww6Scan}{282}
\DefineRemark{totalhostsdnsAAAAwww6Scan6Graph}{404}
\DefineRemark{totalhostsdnsAAAAwww6Scan6Tree}{337}
\DefineRemark{totalhostsdnsAAAAwww6Scan6Tree6Graph}{404}
\DefineRemark{totalhostsgeneratorsonly}{768}
\DefineRemark{totalhostsseedsonly}{5890}

%% file: variables/security.tex
\DefineRemark{amqpcntipv6missaccesscontrol}{750} %
\DefineRemark{ipv4pctmissaccesscontrol}{48} %
\DefineRemark{ipv4pcttls13}{35} %
\DefineRemark{ipv4pcttls13incontentas}{28} %
\DefineRemark{ipv4pcttlsmisconfigured}{1.8} %
\DefineRemark{ipv6pctmissaccesscontrol}{39} %
\DefineRemark{ipv6pcttls13}{63} %
\DefineRemark{ipv6pcttls13incontentas}{63} %
\DefineRemark{ipv6pcttlsmisconfigured}{7.8} %
\DefineRemark{mqttcntipv6missaccesscontrol}{1930} %
\DefineRemark{totalpctaccesscontrolinplace}{52} %
\DefineRemark{totalpctdeprecatedtls}{1.9} %
\DefineRemark{totalpctmissingaccesscontrol}{48} %
\DefineRemark{totalpctmissingsecurity}{94} %

%% file: tables/gen-overview.tex
\begin{tblr}{
  row{1} = {font=\bfseries},
  rowhead = 1,
  column{1,2} = {font=\bfseries},
  cells = {c},
  cell{2}{1} = {c=2,r=2}{},
  cell{4}{1} = {r=15}{},
  cell{4}{2} = {r=10}{},
  cell{14}{2} = {r=5}{},
  hline{2,4} = {-}{},
  hline{14} = {2-5}{},
  colsep = 2.5pt,
  stretch=0,
  cell{3,5,6,7,9,10,11,12,13,14,15,18}{3-Z} = {lightgray}
}
                                                                                                                                                                                                &                                       & Name                                                                        & Year        & Technique                                \\
\begin{sideways}Lists\end{sideways}                                                                                                                                                          &                                       & Song et al.~\cite{song-hitlist-2020}                                        & 2020        & ---                                      \\
                                                                                                                                                                                                &                                       & TUM~\cite{gasser-hitlist-2016, gasser-hitlist-2018, zirngibl-hitlist-2022}  & 2016 / 2022 & ---                                      \\
\begin{tikzpicture} \draw (0.1, 2.6) -- (0,2.6) -- (0,1.25); \node{\begin{sideways}Scanlist Generation\end{sideways}}(0,0); \draw (0.1, -2.6) -- (0, -2.6) -- (0,-1.25); \end{tikzpicture}  & \begin{sideways}Passive\end{sideways} & Ullrich et al.~\cite{ullrich-pattern-2015}                                      & 2015        & Partial Pattern Discovery                \\
                                                                                                                                                                                                &                                       & Entropy/IP~\cite{foremski-entropyip-2016}                                   & 2016        & Bayesian Network (Entropy, Random)       \\
                                                                                                                                                                                                &                                       & 6Gen~\cite{murdock-6gen-2017}                                               & 2017        & High Density Region Fill Up              \\
                                                                                                                                                                                                &                                       & EIP-Generator~\cite{gasser-hitlist-2018}                                    & 2018        & Bayesian Network (Entropy, Complete)     \\
                                                                                                                                                                                                &                                       & IEDC~\cite{zheng-iedc-2020}                                                 & 2020        & High Density Region Fill Up              \\
                                                                                                                                                                                                &                                       & 6GCVAE~\cite{cui-6gcvae-2020}                                               & 2020        & Variational Autoencoder                  \\
                                                                                                                                                                                                &                                       & 6VecLM~\cite{cui-6veclm-2021}                                               & 2020        & Language Model                           \\
                                                                                                                                                                                                &                                       & 6Graph~\cite{yang-6graph-2022}                                              & 2021        & High Density Region Fill Up (Graph)      \\
                                                                                                                                                                                                &                                       & 6GAN~\cite{cui-6gan-2021}                                                   & 2021        & Generative Adversarial Network           \\
                                                                                                                                                                                                &                                       & 6Forest~\cite{yang-6forest-2022}                                            & 2022        & Isolation Forest                         \\
                                                                                                                                                                                                & \begin{sideways}Active\end{sideways}  & 6Tree~\cite{liu-6tree-2019}                                                 & 2019        & Diverse Hierarchical Clustering (DHC)    \\
                                                                                                                                                                                                &                                       & 6Hit~\cite{hou-6hit-2021}                                                   & 2021        & Reinforcement Learning                   \\
                                                                                                                                                                                                &                                       & DET~\cite{song-det-2022}                                                    & 2022        & DHC (incl.\ density and hierarchies)     \\
                                                                                                                                                                                                &                                       & AddrMiner~\cite{song-addrminer-2022}                                        & 2022        & Balanced Spatial Pattern Representation  \\
                                                                                                                                                                                                &                                       & 6Scan~\cite{hou-6scan-2023}                                                 & 2023        & Regional Encoding                                  
\end{tblr}

%% file: tables/dataset.tex
\begin{tabular}{ccccccllcccclccccc} 
\cline{1-7}\cline{9-12}\cline{14-18}
\multicolumn{7}{c}{\textit{Section~\ref{sec:methodology} (Methodology)}}                                                                                                                                                                                                                                                                               &                   & \multicolumn{4}{c}{\textit{Section~\ref{sec:validation} (Validating Responses)}}                                                                                                                                                                                                        & \multicolumn{1}{c}{} & \multicolumn{5}{c}{\textit{Section~\ref{sec:deployments} (Information on Valid Deployments)}}  \\ 
\cline{1-7}\cline{9-12}\cline{14-18}
\multicolumn{2}{c}{\multirow{2}{*}{\textbf{Protocol}}}                                                   & \multirow{2}{*}{\textbf{IP}} & \multirow{2}{*}{\textbf{Port}}  & \multirow{2}{*}{\textbf{Variant}} & \multirow{2}{*}{\begin{tabular}[c]{@{}c@{}}\textbf{Date}\\\textbf{(2023)}\end{tabular}} & \multirow{2}{*}{\textbf{Scanned IPs}}                           & \multirow{2}{*}{} & \multirow{2}{*}{\textbf{Hosts}}                                                                                       & \multirow{2}{*}{\textbf{Transport}}                                                                                            & \multirow{2}{*}{\begin{tabular}[c]{@{}c@{}}\textbf{(D)TLS}\\\textbf{Success}\end{tabular}}   & \multirow{2}{*}{\textbf{Valid}} & \multicolumn{1}{c}{}           & \multicolumn{3}{c}{\textbf{ASes}}                                                                                                                                                                                                                                                                                                & \textbf{Cert. CNs}                                                                                                                        & \multirow{2}{*}{\textbf{Aliasing}}           \\
\multicolumn{2}{c}{}                                                                                     &                              &                                 &                                   &                                                                                         &                                                                 &                   &                                                                                                                       &                                                                                                                                &                                                        &                                   &                                                                  & \textbf{Distinct}                                                                                                                              & \textbf{+IPv6}                                                                        & \textbf{Types (Valid (\%))}\footref{foot:peeringdb}                                       & \textbf{Distinct}                                                                                   &                                              \\ 
\cline{1-7}\cline{9-12}\cline{14-18}                                        
\multirow{13}{*}{\begin{sideways}\textbf{Backend}\end{sideways}} & \multirow{6}{*}{AMQP}                 & \multirow{3}{*}{IPv4}        & \multirow{2}{*}{5672$^\dagger$} & Standard                          & \multirow{2}{*}{\Remark{56725672ipv4amqpdate}}                                          & \multirow{3}{*}{\SI{\Remark{56725672ipv4amqptotalips}}{}}       & \multirow{2}{*}{} & \multirow{2}{*}{\SI{\Remark{56725672ipv4amqptotalhosts}}{}}                                                           & \multirow{2}{*}{\SI{\Remark{56725672ipv4tlsamqptotalconnsuccess}}{}}                                                          & —                                                     & \SI{\Remark{56725672ipv4amqptotalsuccess}}{}      & \multicolumn{1}{c}{}                                & \multirow{3}{*}{\begin{tabular}[c]{@{}c@{}}\SI{\Remark{ipv4amqptotalas}}{}\\(\SI{\Remark{ipv4amqpasperhost}}{}/Valid)\end{tabular}}           & \multirow{3}{*}{—}                                                                    & 1.~\Remark{ipv4amqpastyperank0} (\SI{\Remark{ipv4amqppctastyperank0}}{\percent})          & —                                                                                                                                         & \multirow{3}{*}{—}           \\ 
                                                                    &                                    &                              &                                 & TLS                               &                                                                                         &                                                                 &                   &                                                                                                                       &                                                                                                                               & \SI{\Remark{56725672ipv4tlsamqptotaltlssuccess}}{}    & \SI{\Remark{56725672ipv4tlsamqptotalsuccess}}{}   & \multirow{2}{*}{}                                   &                                                                                                                                               &                                                                                       & 2.~\Remark{ipv4amqpastyperank1} (\SI{\Remark{ipv4amqppctastyperank1}}{\percent})          & \multirow{2}{*}{\begin{tabular}[c]{@{}c@{}}\SI{\Remark{ipv4amqptotalcertcns}}{}\\(\SI{\Remark{tlsipv4amqpcnperhost}}{}/Valid)\end{tabular}}       &          \\ 
\cline{4-6}\cline{9-12}                                     
                                                                    &                                    &                              & 5671$^\ddagger$                 & TLS                               & \Remark{56715671ipv4tlsamqpdate}                                                        &                                                                 &                   & \SI{\Remark{56715671ipv4amqptotalhosts}}{}                                                                            & \SI{\Remark{56715671ipv4tlsamqptotalconnsuccess}}{}                                                                           & \SI{\Remark{56715671ipv4tlsamqptotaltlssuccess}}{}    & \SI{\Remark{56715671ipv4tlsamqptotalsuccess}}{}   &                                                     &                                                                                                                                               &                                                                                       & 3.~\Remark{ipv4amqpastyperank2} (\SI{\Remark{ipv4amqppctastyperank2}}{\percent})          &                                                                                                                                           &            \\ 
\cline{3-7}\cline{9-12}\cline{14-18}                                        
                                                                    &                                    & \multirow{3}{*}{IPv6}        & \multirow{2}{*}{5672$^\dagger$} & Standard                          & \multirow{2}{*}{\Remark{56725672ipv6amqpdate}}                                          & \multirow{2}{*}{\SI{\Remark{56725672ipv6amqptotalips}}{}}       & \multirow{2}{*}{} & \multirow{2}{*}{\SI{\Remark{56725672ipv6amqptotalhosts}}{}}                                                           & \multirow{2}{*}{\SI{\Remark{56725672ipv6tlsamqptotalconnsuccess}}{}}                                                          & —                                                     & \SI{\Remark{56725672ipv6amqptotalsuccess}}{}      & \multicolumn{1}{c}{}                                & \multirow{3}{*}{\begin{tabular}[c]{@{}c@{}}\SI{\Remark{ipv6amqptotalas}}{}\\(\SI{\Remark{ipv6amqpasperhost}}{}/Valid)\end{tabular}}           & \multirow{3}{*}{\SI{\Remark{ipv6amqptotalasv6only}}{}}                                & 1.~\Remark{ipv6amqpastyperank0} (\SI{\Remark{ipv6amqppctastyperank0}}{\percent})          & —                                                                                                                                         & \multirow{3}{*}{\SI{\Remark{ipv6amqptotalaliased}}{}}           \\ 
                                                                    &                                    &                              &                                 & TLS                               &                                                                                         &                                                                 &                   &                                                                                                                       &                                                                                                                               & \SI{\Remark{56725672ipv6tlsamqptotaltlssuccess}}{}    & \SI{\Remark{56725672ipv6tlsamqptotalsuccess}}{}   & \multirow{2}{*}{}                                   &                                                                                                                                               &                                                                                       & 2.~\Remark{ipv6amqpastyperank1} (\SI{\Remark{ipv6amqppctastyperank1}}{\percent})          & \multirow{2}{*}{\begin{tabular}[c]{@{}c@{}}\SI{\Remark{ipv6amqptotalcertcns}}{}\\(\SI{\Remark{tlsipv6amqpcnperhost}}{}/Valid)\end{tabular}}       &          \\ 
\cline{4-7}\cline{9-12}                                     
                                                                    &                                    &                              & 5671$^\ddagger$                 & TLS                               & \Remark{56715671ipv6tlsamqpdate}                                                        & \SI{\Remark{56715671ipv6tlsamqptotalips}}{}                     &                   & \SI{\Remark{56715671ipv6amqptotalhosts}}{}                                                                            & \SI{\Remark{56715671ipv6tlsamqptotalconnsuccess}}{}                                                                            & \SI{\Remark{56715671ipv6tlsamqptotaltlssuccess}}{}    & \SI{\Remark{56715671ipv6tlsamqptotalsuccess}}{}   &                                                     &                                                                                                                                              &                                                                                       & 3.~\Remark{ipv6amqpastyperank2} (\SI{\Remark{ipv6amqppctastyperank2}}{\percent})          &                                                                                                                                           &            \\ 
\cline{2-7}\cline{9-12}\cline{14-18}                                        
                                                                    & \multirow{6}{*}{MQTT}              & \multirow{3}{*}{IPv4}        & \multirow{2}{*}{1883$^\dagger$} & Standard                          & \multirow{2}{*}{\Remark{18831883ipv4mqttdate}}                                          & \multirow{3}{*}{\SI{\Remark{18831883ipv4mqtttotalips}}{}}       & \multirow{2}{*}{} & \multirow{2}{*}{\SI{\Remark{18831883ipv4mqtttotalhosts}}{}}                                                           & \multirow{2}{*}{\SI{\Remark{18831883ipv4tlsmqtttotalconnsuccess}}{}}                                                          & —                                                     & \SI{\Remark{18831883ipv4mqtttotalsuccess}}{}      & \multicolumn{1}{c}{}                                & \multirow{3}{*}{\begin{tabular}[c]{@{}c@{}}\SI{\Remark{ipv4mqtttotalas}}{}\\(\SI{\Remark{ipv4mqttasperhost}}{}/Valid)\end{tabular}}           & \multirow{3}{*}{—}                                                                    & 1.~\Remark{ipv4mqttastyperank0} (\SI{\Remark{ipv4mqttpctastyperank0}}{\percent})          & —                                                                                                                                         & \multirow{3}{*}{—}           \\ 
                                                                    &                                    &                              &                                 & TLS                               &                                                                                         &                                                                 &                   &                                                                                                                       &                                                                                                                               & \SI{\Remark{18831883ipv4tlsmqtttotaltlssuccess}}{}    & \SI{\Remark{18831883ipv4tlsmqtttotalsuccess}}{}   & \multirow{2}{*}{}                                   &                                                                                                                                               &                                                                                       & 2.~\Remark{ipv4mqttastyperank1} (\SI{\Remark{ipv4mqttpctastyperank1}}{\percent})          & \multirow{2}{*}{\begin{tabular}[c]{@{}c@{}}\SI{\Remark{ipv4mqtttotalcertcns}}{}\\(\SI{\Remark{tlsipv4mqttcnperhost}}{}/Valid)\end{tabular}}       &          \\ 
\cline{4-6}\cline{9-12}                                     
                                                                    &                                    &                              & 8883$^\ddagger$                 & TLS                               & \Remark{88838883ipv4tlsmqttdate}                                                        &                                                                 &                   & \SI{\Remark{88838883ipv4mqtttotalhosts}}{}                                                                            & \SI{\Remark{88838883ipv4tlsmqtttotalconnsuccess}}{}                                                                            & \SI{\Remark{88838883ipv4tlsmqtttotaltlssuccess}}{}    & \SI{\Remark{88838883ipv4tlsmqtttotalsuccess}}{}   &                                                     &                                                                                                                                              &                                                                                       & 3.~\Remark{ipv4mqttastyperank2} (\SI{\Remark{ipv4mqttpctastyperank2}}{\percent})          &                                                                                                                                           &          \\ 
\cline{3-7}\cline{9-12}\cline{14-18}                                        
                                                                    &                                    & \multirow{3}{*}{IPv6}        & \multirow{2}{*}{1883$^\dagger$} & Standard                          & \multirow{2}{*}{\Remark{18831883ipv6mqttdate}}                                          & \multirow{2}{*}{\SI{\Remark{18831883ipv6mqtttotalips}}{}}       & \multirow{2}{*}{} & \multirow{2}{*}{\SI{\Remark{18831883ipv6mqtttotalhosts}}{}}                                                           & \multirow{2}{*}{\SI{\Remark{18831883ipv6tlsmqtttotalconnsuccess}}{}}                                                          & —                                                     & \SI{\Remark{18831883ipv6mqtttotalsuccess}}{}      & \multicolumn{1}{c}{}                                & \multirow{3}{*}{\begin{tabular}[c]{@{}c@{}}\SI{\Remark{ipv6mqtttotalas}}{}\\(\SI{\Remark{ipv6mqttasperhost}}{}/Valid)\end{tabular}}           & \multirow{3}{*}{\SI{\Remark{ipv6mqtttotalasv6only}}{}}                                & 1.~\Remark{ipv6mqttastyperank0} (\SI{\Remark{ipv6mqttpctastyperank0}}{\percent})          & —                                                                                                                                         & \multirow{3}{*}{\SI{\Remark{ipv6mqtttotalaliased}}{}}          \\ 
                                                                    &                                    &                              &                                 & TLS                               &                                                                                         &                                                                 &                   &                                                                                                                       &                                                                                                                               & \SI{\Remark{18831883ipv6tlsmqtttotaltlssuccess}}{}    & \SI{\Remark{18831883ipv6tlsmqtttotalsuccess}}{}   & \multirow{2}{*}{}                                   &                                                                                                                                               &                                                                                       & 2.~\Remark{ipv6mqttastyperank1} (\SI{\Remark{ipv6mqttpctastyperank1}}{\percent})          & \multirow{2}{*}{\begin{tabular}[c]{@{}c@{}}\SI{\Remark{ipv6mqtttotalcertcns}}{}\\(\SI{\Remark{tlsipv6mqttcnperhost}}{}/Valid)\end{tabular}}       &          \\ 
\cline{4-7}\cline{9-12}      
                                                                    &                                    &                              & 8883$^\ddagger$                 & TLS                               & \Remark{88838883ipv6tlsmqttdate}                                                        & \SI{\Remark{88838883ipv6tlsmqtttotalips}}{}                     &                   & \SI{\Remark{88838883ipv6mqtttotalhosts}}{}                                                                            & \SI{\Remark{88838883ipv6tlsmqtttotalconnsuccess}}{}                                                                            & \SI{\Remark{88838883ipv6tlsmqtttotaltlssuccess}}{}    & \SI{\Remark{88838883ipv6tlsmqtttotalsuccess}}{}   &                                                     &                                                                                                                                              &                                                                                       & 3.~\Remark{ipv6mqttastyperank2} (\SI{\Remark{ipv6mqttpctastyperank2}}{\percent})          &                                                                                                                                           &            \\ 
\cline{1-7}\cline{9-12}\cline{14-18}                                        
\multirow{13}{*}{\begin{sideways}\textbf{Device}\end{sideways}}     & \multirow{6}{*}{OPC UA}            & \multirow{3}{*}{IPv4}        & \multirow{2}{*}{4840$^\dagger$} & Standard                          & \multirow{2}{*}{\Remark{48404840ipv4opcuadate}}                                          & \multirow{3}{*}{\SI{\Remark{48404840ipv4opcuatotalips}}{}}       & \multirow{2}{*}{} & \multirow{2}{*}{\SI{\Remark{48404840ipv4opcuatotalhosts}}{}}                                                        & \multirow{2}{*}{\SI{\Remark{48404840ipv4tlsopcuatotalconnsuccess}}{}}                                                         & —                                                      & \SI{\Remark{48404840ipv4opcuatotalsuccess}}{}      & \multicolumn{1}{c}{}                               & \multirow{3}{*}{\begin{tabular}[c]{@{}c@{}}\SI{\Remark{ipv4opcuatotalas}}{}\\(\SI{\Remark{ipv4opcuaasperhost}}{}/Valid)\end{tabular}}        & \multirow{3}{*}{—}                                                                    & 1.~\Remark{ipv4opcuaastyperank0} (\SI{\Remark{ipv4opcuapctastyperank0}}{\percent})        & \multirow{3}{*}{\begin{tabular}[c]{@{}c@{}}\SI{\Remark{ipv4opcuatotalcertcns}}{}\\(\SI{\Remark{ipv4opcuacnperhost}}{}/Valid)\end{tabular}}                     & \multirow{3}{*}{—}          \\ 
                                                                    &                                    &                              &                                 & TLS                               &                                                                                         &                                                                 &                   &                                                                                                                       &                                                                                                                               & \SI{\Remark{48404840ipv4tlsopcuatotaltlssuccess}}{}    & \SI{\Remark{48404840ipv4tlsopcuatotalsuccess}}{}   & \multirow{2}{*}{}                                 &                                                                                                                                               &                                                                                       & 2.~\Remark{ipv4opcuaastyperank1} (\SI{\Remark{ipv4opcuapctastyperank1}}{\percent})        &       &           \\ 
\cline{4-6}\cline{9-12}                                     
                                                                    &                                    &                              & 4843$^\ddagger$                 & TLS                               & \Remark{48434843ipv4tlshttpdate}                                                        &                                                                 &                   & \SI{\Remark{48434843ipv4httptotalhosts}}{}                                                                            & \SI{\Remark{48434843ipv4tlshttptotalconnsuccess}}{}                                                                            & \SI{\Remark{48434843ipv4tlshttptotaltlssuccess}}{}    & \SI{\Remark{48434843ipv4tlshttptotalsuccess}}{}   &                                                     &                                                                                                                                              &                                                                                       & 3.~\Remark{ipv4opcuaastyperank2} (\SI{\Remark{ipv4opcuapctastyperank2}}{\percent})        &                                                                                                                                           &            \\ 
\cline{3-7}\cline{9-12}\cline{14-18}                                        
                                                                    &                                    & \multirow{3}{*}{IPv6}        & \multirow{2}{*}{4840$^\dagger$} & Standard                          & \multirow{2}{*}{\Remark{48404840ipv6opcuadate}}                                          & \multirow{2}{*}{\SI{\Remark{48404840ipv6opcuatotalips}}{}}       & \multirow{2}{*}{} & \multirow{2}{*}{\SI{\Remark{48404840ipv6opcuatotalhosts}}{}}                                                        & \multirow{2}{*}{\SI{\Remark{48404840ipv6tlsopcuatotalconnsuccess}}{}}                                                         & —                                                       & \SI{\Remark{48404840ipv6opcuatotalsuccess}}{}      & \multicolumn{1}{c}{}                               & \multirow{3}{*}{\begin{tabular}[c]{@{}c@{}}\SI{\Remark{ipv6opcuatotalas}}{}\\(\SI{\Remark{ipv6opcuaasperhost}}{}/Valid)\end{tabular}}       & \multirow{3}{*}{\SI{\Remark{ipv6opcuatotalasv6only}}{}}                             & 1.~\Remark{ipv6opcuaastyperank0} (\SI{\Remark{ipv6opcuapctastyperank0}}{\percent})          & \multirow{3}{*}{\begin{tabular}[c]{@{}c@{}}\SI{\Remark{ipv6opcuatotalcertcns}}{}\\(\SI{\Remark{tlsipv6opcuacnperhost}}{}/Valid)\end{tabular}}                     & \multirow{3}{*}{\SI{\Remark{ipv6opcuatotalaliased}}{}}            \\ 
                                                                    &                                    &                              &                                 & TLS                               &                                                                                         &                                                                 &                   &                                                                                                                       &                                                                                                                               & \SI{\Remark{48404840ipv6tlsopcuatotaltlssuccess}}{}    & \SI{\Remark{48404840ipv6tlsopcuatotalsuccess}}{}   & \multirow{2}{*}{}                                 &                                                                                                                                               &                                                                                       & 2.~\Remark{ipv6opcuaastyperank1} (\SI{\Remark{ipv6opcuapctastyperank1}}{\percent})        &       &           \\ 
\cline{4-7}\cline{9-12}                                     
                                                                    &                                    &                              & 4843$^\ddagger$                 & TLS                               & \Remark{48434843ipv6tlshttpdate}                                                        & \SI{\Remark{48434843ipv6tlshttptotalips}}{}                     &                   & \SI{\Remark{48434843ipv6httptotalhosts}}{}                                                                            & \SI{\Remark{48434843ipv6tlshttptotalconnsuccess}}{}                                                                            & \SI{\Remark{48434843ipv6tlshttptotaltlssuccess}}{}    & \SI{\Remark{48434843ipv6tlshttptotalsuccess}}{}   &                                                     &                                                                                                                                              &                                                                                       & 3.~\Remark{ipv6opcuaastyperank2} (\SI{\Remark{ipv6opcuapctastyperank2}}{\percent})        &                                                                                                                                           &            \\ 
\cline{2-7}\cline{9-12}\cline{14-18}                                        
                                                                    & \multirow{6}{*}{CoAP}              & \multirow{3}{*}{IPv4}        & \multirow{2}{*}{5683$^\dagger$} & Standard                          & \multirow{2}{*}{\Remark{56835683ipv4coapdate}}                                          & \multirow{3}{*}{\SI{\Remark{56835683ipv4coaptotalips}}{}}       & \multirow{2}{*}{} & \multirow{2}{*}{\SI{\Remark{56835683ipv4coaptotalhosts}}{}}                                                           & \multirow{2}{*}{\SI{\Remark{56835683ipv4tlscoaptotalconnsuccess}}{}}                                                          & —                                                     & \SI{\Remark{56835683ipv4coaptotalsuccess}}{}      & \multicolumn{1}{c}{}                                & \multirow{3}{*}{\begin{tabular}[c]{@{}c@{}}\SI{\Remark{ipv4coaptotalas}}{}\\(\SI{\Remark{ipv4coapasperhost}}{}/Valid)\end{tabular}}           & \multirow{3}{*}{—}                                                                    & 1.~\Remark{ipv4coapastyperank0} (\SI{\Remark{ipv4coappctastyperank0}}{\percent})          & —                                                                                                                                         & \multirow{3}{*}{—}           \\ 
                                                                    &                                    &                              &                                 & DTLS                              &                                                                                         &                                                                 &                   &                                                                                                                       &                                                                                                                               & \SI{\Remark{56835683ipv4tlscoaptotaltlssuccess}}{}    & \SI{\Remark{56835683ipv4tlscoaptotalsuccess}}{}   & \multirow{2}{*}{}                                   &                                                                                                                                               &                                                                                       & 2.~\Remark{ipv4coapastyperank1} (\SI{\Remark{ipv4coappctastyperank1}}{\percent})          & \multirow{2}{*}{\begin{tabular}[c]{@{}c@{}}\SI{\Remark{ipv4coaptotalcertcns}}{}\\(\SI{\Remark{tlsipv4coapcnperhost}}{}/Valid)\end{tabular}}       &          \\ 
\cline{4-6}\cline{9-12}                                     
                                                                    &                                    &                              & 5684$^\ddagger$                 & DTLS                              & \Remark{56845684ipv4tlscoapdate}                                                        &                                                                 &                   & \SI{\Remark{56845684ipv4coaptotalhosts}}{}                                                                            & \SI{\Remark{56845684ipv4tlscoaptotalconnsuccess}}{}                                                                            & \SI{\Remark{56845684ipv4tlscoaptotaltlssuccess}}{}    & \SI{\Remark{56845684ipv4tlscoaptotalsuccess}}{}   &                                                     &                                                                                                                                              &                                                                                       & 3.~\Remark{ipv4coapastyperank2} (\SI{\Remark{ipv4coappctastyperank2}}{\percent})          &                                                                                                                                           &          \\ 
\cline{3-7}\cline{9-12}\cline{14-18}                                        
                                                                    &                                    & \multirow{3}{*}{IPv6}        & \multirow{2}{*}{5683$^\dagger$} & Standard                          & \multirow{2}{*}{06-24}                                                                  & \multirow{2}{*}{\SI{2219964687}{}}                              & \multirow{2}{*}{} & \multirow{2}{*}{\SI{0}{}}                                                                                             & \multirow{2}{*}{\SI{0}{}}                                                                                                     & —                                                     & \SI{0}{}                                           & \multicolumn{1}{c}{}                                & \multirow{3}{*}{\begin{tabular}[c]{@{}c@{}}\SI{\Remark{ipv6coaptotalas}}{}\\(\SI{\Remark{ipv6coapasperhost}}{}/Valid)\end{tabular}}           & \multirow{3}{*}{\SI{\Remark{ipv6coaptotalasv6only}}{}}                              & 1.~\Remark{ipv6coapastyperank0} (\SI{\Remark{ipv6coappctastyperank0}}{\percent})          & —                                                                                                                                         & \multirow{3}{*}{\SI{\Remark{ipv6coaptotalaliased}}{}}          \\ 
                                                                    &                                    &                              &                                 & DTLS                              &                                                                                         &                                                                 &                   &                                                                                                                       &                                                                                                                               & \SI{0}{}                                              & \SI{0}{}                                           & \multirow{2}{*}{}                                   &                                                                                                                                               &                                                                                       & 2.~\Remark{ipv6coapastyperank1} (\SI{\Remark{ipv6coappctastyperank1}}{\percent})         & \multirow{2}{*}{\begin{tabular}[c]{@{}c@{}}\SI{\Remark{ipv6coaptotalcertcns}}{}\\(\SI{\Remark{tlsipv6coapcnperhost}}{}/Valid)\end{tabular}}       &          \\ 
\cline{4-7}\cline{9-12}      
                                                                    &                                    &                              & 5684$^\ddagger$                 & DTLS                              & \Remark{56845684ipv6tlscoapdate}                                                        & \SI{\Remark{56845684ipv6tlscoaptotalips}}{}                     &                   & \SI{\Remark{56845684ipv6coaptotalhosts}}{}                                                                            & \SI{\Remark{56845684ipv6tlscoaptotalconnsuccess}}{}                                                                            & \SI{\Remark{56845684ipv6tlscoaptotaltlssuccess}}{}    & \SI{\Remark{56845684ipv6tlscoaptotalsuccess}}{}   &                                                     &                                                                                                                                              &                                                                                       & 3.~\Remark{ipv6coapastyperank2} (\SI{\Remark{ipv6coappctastyperank2}}{\percent})          &    &            \\ 
\cline{1-7}\cline{9-12}\cline{14-18}                                        
\end{tabular}                                       